\documentclass[12pt]{article}

\usepackage{newtxtext,newtxmath}

\usepackage{graphicx}

\usepackage[letterpaper,margin=1in]{geometry}

\renewenvironment{abstract}
	{\quotation}
	{\endquotation}

\date{}

\makeatletter
\renewcommand{\fnum@figure}{\textbf{Figure \thefigure}}
\renewcommand{\fnum@table}{\textbf{Table \thetable}}
\makeatother

\usepackage{scicite}

\usepackage{url}

\def\scititle{
	Is Seismic Forecasting Possible \\ with Physics-based AI?
}
\title{\bfseries \boldmath \scititle}

\author{
    V.~Keane$^{1\ast}$,
	T.~Poulet$^{2}$,
    E.~Veveakis$^{1}$\and
	\small$^{1}$Department of Civil and Environmental Engineering, Duke University, Durham NC, USA.\and
	\small$^{2}$ Earth Process Models, CSIRO, Kensington WA, Australia.\and
	\small$^\ast$Corresponding author. Email: victoria.keane@duke.edu\and
}

\begin{document} 

\maketitle

\begin{abstract} \bfseries \boldmath
Slow slip events within subduction zones offer a unique window into earthquake prediction. The subducting plate drives dehydration reactions in the fault, causing cyclical slip and observable surface displacements. Earthquake footprints can then be identified in these displacement series through coupling the multi-physics governing the subduction process with regional seismic activity. However, data noise and traditional filtering methods obscure the underlying mechanisms. Here, we alleviate this constraint with our physics-based attractor. By accounting for the physics of subduction paired with AI-assisted manifold detection, we are able to predict an earthquake in New Zealand's Hikurangi trench one week early. Additionally, predictability limits extend to 5-6 weeks with decadal repeatability, pointing to the fundamental determinism of the suggested mechanism through which physics-based seismic forecasting is possible.

\end{abstract}

\noindent
Slow Slip Events (SSEs) are episodic releases of tectonic energy that occur over days to weeks, in contrast to their quick-slip counterparts that rupture in seconds. These events are widely observed in subduction zones and often exhibit periodic or quasi-periodic recurrence, making them a useful system for investigating fault dynamics \cite{dragert, wallace2004, voss, kodaira, hirose1999, obara2002, gualandi2025}. While these sequences often appear erratic, recent work suggests they represent a form of predictable chaos, where the underlying governing laws are deterministic despite their apparent complexity \cite{voss, Gualandietal2020, slowslip}. 

Historically, deterministic seismic hazard forecasting has remained out of reach.  The unpredictable cascade of fast-slip earthquakes relies on initial stress and frictional conditions that cannot be resolved at depth, forcing the field to rely heavily on probabilistic models rather than deterministic physical laws \cite{geller1997, jordan2006, bayona2026}.  However, the prolonged, continuous nature of SSE fault movements provides a new window into these hidden mechanics, offering the steady, observable surface displacements necessary to invert for the driving physical processes.

In recent decades, SSEs have been extensively studied thanks to the widespread deployment of Global Navigation Satellite System (GNSS) stations \cite{part3, wallace2004, kodaira, Hulbertetal2020, dragert, hirose1999, gualandi2025}. These sensors provide the surface displacement data necessary to investigate the physical mechanisms driving SSEs, allowing for the derivation of physics-based models. We adopt a nonlinear coupled system of partial differential equations (PDEs) modeling the evolution of pore pressure and temperature across a shear zone (Eq.~\ref{eq:full_system}, Supplementary Text) \cite{part1, part2, part3}. Physically, the processes occurring correspond to the thermally activated dehydration reaction of serpentine minerals which has been identified as a dominant driver of instability within subduction zones\cite{dobson, part3}. 

The map and cross section of Cascadia (Fig.~\ref{fig:CASC_map}A and B) reveal the distinction between locked and slipping zones of faults.  At the deeper, higher temperature SSE location on the fault interface, the system reaches a critical bifurcation dependent on temperature competing with pressure, producing consistent cyclical signals.  The core of this mechanism is the competition between fluid generation and diffusion. As the overriding plate creeps, frictional heating and deformation increase temperatures within the shear zone. Once a critical threshold is met, an endothermic reaction is triggered, releasing fluid from the solid skeleton of the rock, increasing pore pressure, decreasing effective stress of the interface and causing it to slip \cite{part1, part2, part3, dobson}. 

\subsection*{Physics-Based Phase Space Reconstruction}

In this work we are using a thermo-poro-mechanical (multi-physics) framework as a physics-based filter to guide our inversion and forecast. The mathematical details of the framework can be found in \cite{part1,part2,part3} and abbreviated in the methods section. 

In Cascadia, our physics-based filter isolates the almost perfectly periodic displacement signal from the ALBH sensor (Fig.~\ref{fig:CASC_map}C). Using Taken's Embedding Theorem \cite{takens1981}, the filtered time series can be used to reconstruct the higher-dimensional phase spaces. This method derives delay vectors from the single-valued displacement time series to extract system dynamics that may otherwise be obscured by noise or sensor quality (Eq.~\ref{eq:delay_vectors}) \cite{takens1981, abarbanel1993, methods}.  Crucially, this mathematically derived 3D attractor directly maps to the three governing state variables of our multi-physics framework: temperature, pore pressure, and effective stress. Compared to reconstructions derived from the raw data and simple Savitzky-Golay filter \cite{savgol1964} (Fig.~\ref{fig:CASC_map}D and E), the physics filter (Fig.~\ref{fig:CASC_map}F) yields a remarkably clean, low-dimensional phase space.  

Data filtering is arguably the most critical step in phase space reconstruction. Cascadia provides a pristine test location for our physics filter due to the lack of high seismic activity (earthquakes > magnitude 5.0). In these simpler systems, the elastically deforming overriding plate acts primarily as a critically charged spring, gradually relaxing the system back into its fundamental mode \cite{part1, part2, part3, slowslip}. This aligns with observations of an exponential build-up in seismic energy prior to slip, suggesting that Cascadia SSEs undergo a months-long nucleation phase rather than a sudden stochastic trigger \cite{Hulbertetal2020}. 

In contrast, more complex subduction zone systems, such as the Hikurangi trench, exhibit SSEs that oscillate between two distinct modes (Fig.~\ref{fig:NZ_map}B), suggesting much richer underlying dynamics \cite{part1, slowslip, wallace2004}.  New Zealand's North Island is a prime area of study due to its proximity to the active Hikurangi subduction margin and the availability of high-quality GNSS sensor data dating back to 2002 \cite{wallace2004, geonet}. Fig.~\ref{fig:NZ_map}A shows a schematic of the North Island with shaded regions corresponding to the active SSE and locked regions, alongside the GNSS stations utilized in this study. The core of the SSE signal from each station is isolated through a global detrending using the linear, first-order polynomial fit to remove background tectonic drift \cite{methods}. Environmental noise, primarily from hydrological and atmospheric loading, is addressed by removing annual and semi-annual seasonal signals \cite{methods, liu2020}. 

We manually overlay the physics filter on the GISB cleaned displacement data (Fig.~\ref{fig:NZ_map}B).  In this more complex system, the physics filter shifts between two distinct modes (red and black), assumed to be a physical manifestation of the competing deep and shallow SSE occurring on the island \cite{part1, slowslip}. These two modes are derived from the same coupled PDE system with exactly the same parameter values (see Table 3 on page 4618 in \cite{part3}), differing only on the value of the boundary (regional) stresses that the overriding plate applies to the subduction interface \cite{part1, part2, part3}. 

The red mode (lower stresses) corresponds to shallower, lower-energy processes leading to larger displacement values and longer recurrence periods \cite{part1, part2, part3, slowslip}. Conversely, the black mode (higher stresses) is roughly the same as ALBH, Fig. \ref{fig:CASC_map}C, pointing to its fundamental nature as the mode of oscillation in subduction environments. This mode experiences more frequent, milder events typical of deeper SSE sources (Supplementary Text). 

Regular displacement time series, like the one in Cascadia, can be modeled using a constant in time value of the regional stresses. On the other hand, irregular displacement time series, like the one in GISB, require additional information on the nature of the transition between high and low regional stresses in subducting environments. We posit here that transitions from red to black track natural loading on the overriding plate, while a transition from black back to red requires external energy (such as a nearby seismic event) to release accumulated stress and return the system to its fundamental mode. This observation allows us to identify earthquake events on irregular displacement time series like the one in GISB.

To do so, we once more find the phase space reconstruction of the manually clipped physical modes to be remarkably clean and planar (Fig.~\ref{fig:NZ_map}E). While reconstructions from raw GNSS residuals (Fig.~\ref{fig:NZ_map}C) and standard Savitzky-Golay (Fig.~\ref{fig:NZ_map}D) filters remain convoluted, failing to fully resolve the signal from the stochastic jitter, our physics-based clipping isolates the low-dimensional manifold governing the system’s state variables. This reconstruction implies that the seemingly chaotic transients of the Hikurangi subduction zone are not purely stochastic, but are governed by a deterministic interplay between internal fault physics and external forcing.

\subsection*{Automated Manifold Optimization in Complex Margins}

To transition from manual to automated clipping, we interpret the quasi-periodicity of Hikurangi SSEs as a dynamical coupling governed by the evolution of the regional stresses. We introduce a reduced-order modal evolution defined by a first-order relaxation Ordinary Differential Equation (ODE), Equation \ref{eq:relax_ode} in the methods section.  

This equation features a modal weight order parameter $\phi$, and a characteristic relaxation time of the fault interface, $\tau$.  In this framework, the system naturally evolves under tectonic loading from the red mode ($\phi=0$) toward the black mode ($\phi=1$), resetting at discrete event dates, representing abrupt stress drops on the interface. These discrete events are derived using a multi-stage saliency filter on the United States Geological Survey (USGS) seismic catalog \cite{usgs, methods} (table~\ref{tab:event_dates}). Model parameters were estimated through an ensemble optimization using Differential Evolution \cite{storn}, employing 15 independent trajectories per station to produce a probabilistic envelope and final mean fits \cite{methods}. 

For the GISB station, the automated event selection identifies the four black-to-red transitions as significant earthquake events, alongside five other nearby discrete events.  The disagreement of the filter in the two dashed green boxes (Fig.~\ref{fig:NZmap_events}) highlights the strict physical discipline of the model. The system correctly identifies non-tectonic sensor perturbations and data sparsity as physical anomalies, refusing to produce an overfitted solution that violates the governing thermo-poro-mechanical laws.  Furthermore, during quiescent periods, the model correctly identifies the absence of a physical bifurcation across the network, signaling a stable non-nucleation state rather than projecting spurious slip trajectories.

Using these optimized mean fits, we again employ Taken's Embedding theorem to reconstruct the phase spaces \cite{methods}. By leveraging the bounded geometry of these resolved manifolds, we implement a station-adaptive K-Nearest Neighbors (KNN) algorithm to predict the future evolution of the fault interface \cite{fan2021, methods}.  Rather than relying on rigid statistical extrapolation, KNN leverages the intrinsic determinism of the system, projecting future states by identifying localized historical analogs and tracking their subsequent trajectories.

\subsection*{Forecasting Results and Discussion}

\begin{figure}
	\centering
  \includegraphics[width=1\textwidth]{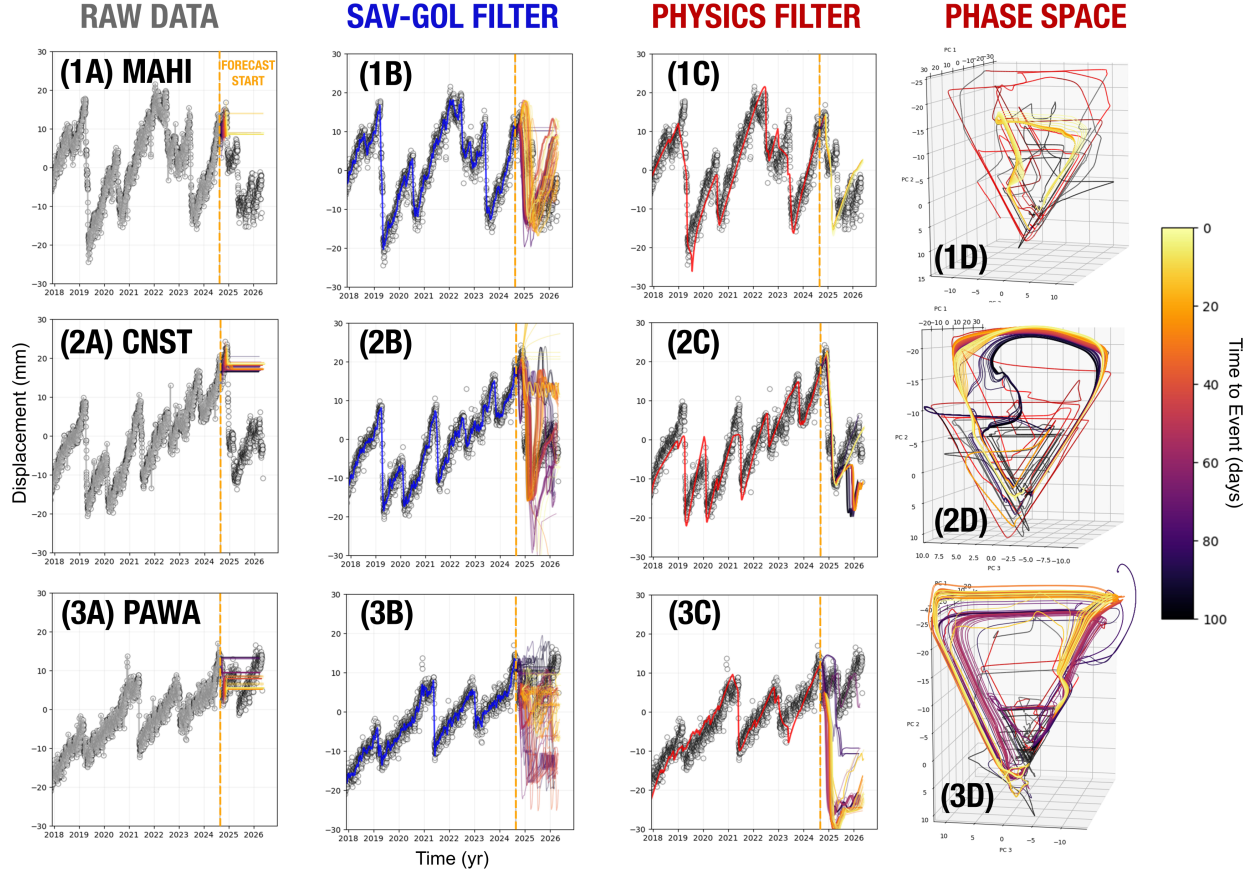} 
	\caption{\textbf{Physics-based phase space reconstruction enables deterministic forecasting of slow slip dynamics.} Displacement forecasts for the 2024 M5.8 event extending for 1.5 years from stations MAHI (\textbf{1}), CNST (\textbf{2}), and PAWA (\textbf{3}).  Forecasts are initialized at daily intervals starting 100 days prior to the event (yellow dashed line), with initialization times color-coded from dark blue (100 days prior) to light yellow (day of event).  (\textbf{A to C}) Time series forecasts from the raw data, Savitzky-Golay filter, and physics filter, respectively.  K-nearest neighbors (KNN) algorithm was trained on the corresponding phase spaces and used to construct the forecasts.  The physics filter (\textbf{C}) achieves remarkable convergence compared to the raw data (\textbf{A}) and Savitzky-Golay filter (\textbf{B}).  Stations MAHI and CNST show highly consistent, tightly bounded predictions.  The broader predictive variance at PAWA, a more distant southern station, physically reflects that this localized fault segment was not yet critically stressed to break.  (\textbf{D}) Physics-based phase space reconstructions overlaid with the forecast trajectories.  The KNN algorithm relies on the system's underlying determinism, allowing it to accurately project future states by following established historical trajectories along the low-dimensional manifold.
  }
	\label{fig:forecasts} 
\end{figure}

We focus our forecasting evaluation on the top three stations, achieving the highest $R^2$ values from the optimized fitting, for the final 2024 M5.8 earthquake (Fig.~\ref{fig:forecasts}, table~\ref{tab:fit_accuracy}).  Station MAHI achieved remarkable convergence and accuracy in its prediction.  Beginning 100 days before the event and extending predictions for 1.5 years, the trajectory not only accurately captures the seismic event but also successfully tracks the reloading rate of the subsequent cycle (Fig.~\ref{fig:forecasts}1C).  In contrast, the raw data forecasts completely break down (Fig.~\ref{fig:forecasts}1A), and the Savitzky-Golay filter exhibits high disagreement and overestimates displacement following the subsequent jump (Fig.~\ref{fig:forecasts}1B). Compared to the tangled phase spaces of the raw data and Savitzky-Golay filters, the physics filter trajectories navigate the bounded, low dimensional paths with significantly greater ease and convergence (Fig.~\ref{fig:forecasts}1-3D).  

\begin{figure}
	\centering
	\includegraphics[width=1\textwidth]{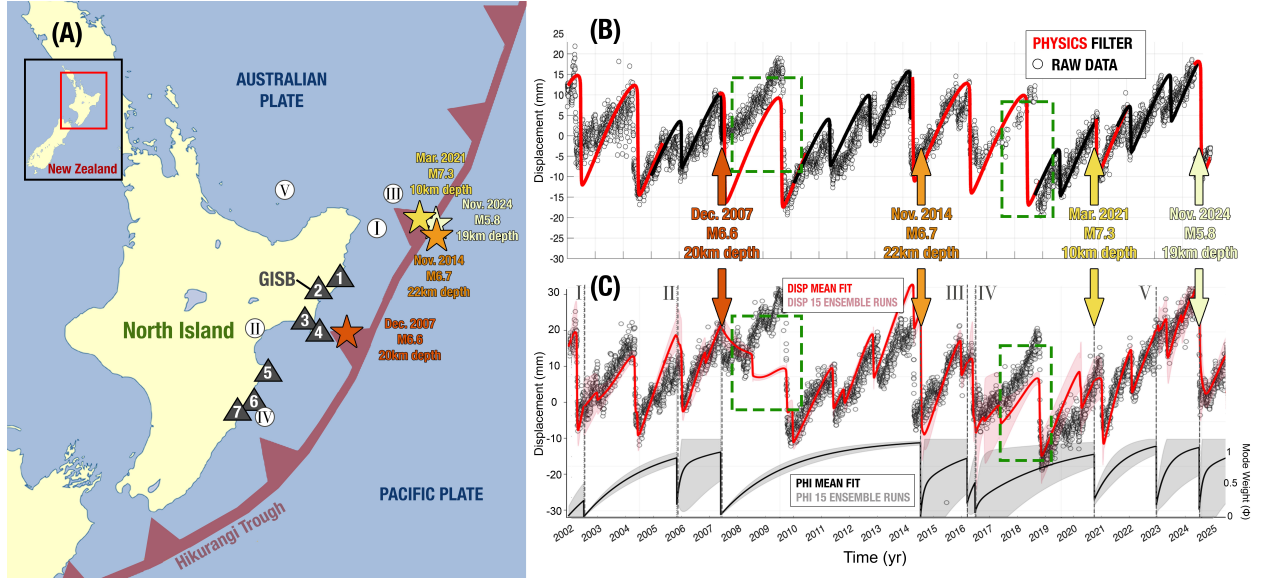} 
	\caption{\textbf{Automated physics-based optimization of Hikurangi slow slip dynamics.}
  (\textbf{A}) Map of the Hikurangi margin detailing discrete seismic events identified by the optimization pipeline.  Stars denote events that align with manual selections, while circles mark additional events (I to V) captured by the multi-stage saliency filter (I. Feb. 2003, M5.5, 58.9 km depth; II. Jul. 2006, M5.3, 30 km depth; III. Sept. 2016, M7.0, 19.0 km depth; IV. Nov. 2016, M5.9, 9.1 km depth; V. May 2023, M5.6, 186.0 km depth). (\textbf{B and C}) GISB displacement data overlaid with the manually clipped physics filter and the automated optimization, respectively.  The automated fit (\textbf{C}) displays the mean trajectory (red) and a 15-run probabilistic ensemble mask (light red). The black-to-red modal transitions from the manual clipping, which require external excitation, align perfectly with the four star-marked events, while the remaining five selected events (I-V) are indicated by dotted lines.  The optimizer's failure to track data within regions of sensor jumps and data sparsity (green dashed boxes) highlights its strict adherence to physical laws rather than statistical overfitting.  (\textbf{C, bottom}) Evolution of the modal weight parameter ($\phi$). Total system displacement is defined by $u_{total} = \phi u_{black} + (1-\phi) u_{red}$, where $\phi$ relaxes toward $1$ under natural tectonic loading and resets toward $0$ in response to the discrete seismic triggers.} 
  \label{fig:NZmap_events}
\end{figure}

Station CNST shows slightly more variance for the physics filter, but still substantially outperforms the raw data and Savitzky-Golay methods (Fig.~\ref{fig:forecasts}2).  Station PAWA exhibits a higher deviation between the predicted and observed displacement magnitudes; however, this divergence provides critical insight into the spatial limits of the macroscopic physical system.  The model accurately captured the temporal onset of the regional instability.  Yet, because PAWA is located significantly south of the event epicenter (Fig.~\ref{fig:NZmap_events}A), the localized magnitude of the slip was physically hindered.  The true time series indicates that the southern portion of the fault was prevented from fully slipping, likely due to a strong local asperity or the rupture running out of fracture energy.  This spatial variance can be addressed in future studies by incorporating higher degrees of freedom for localized rupture and implementing full network station coordinates to gain a spatiotemporal understanding of the entire fault system.

\begin{figure}
	\centering
  \includegraphics[width=1\textwidth]{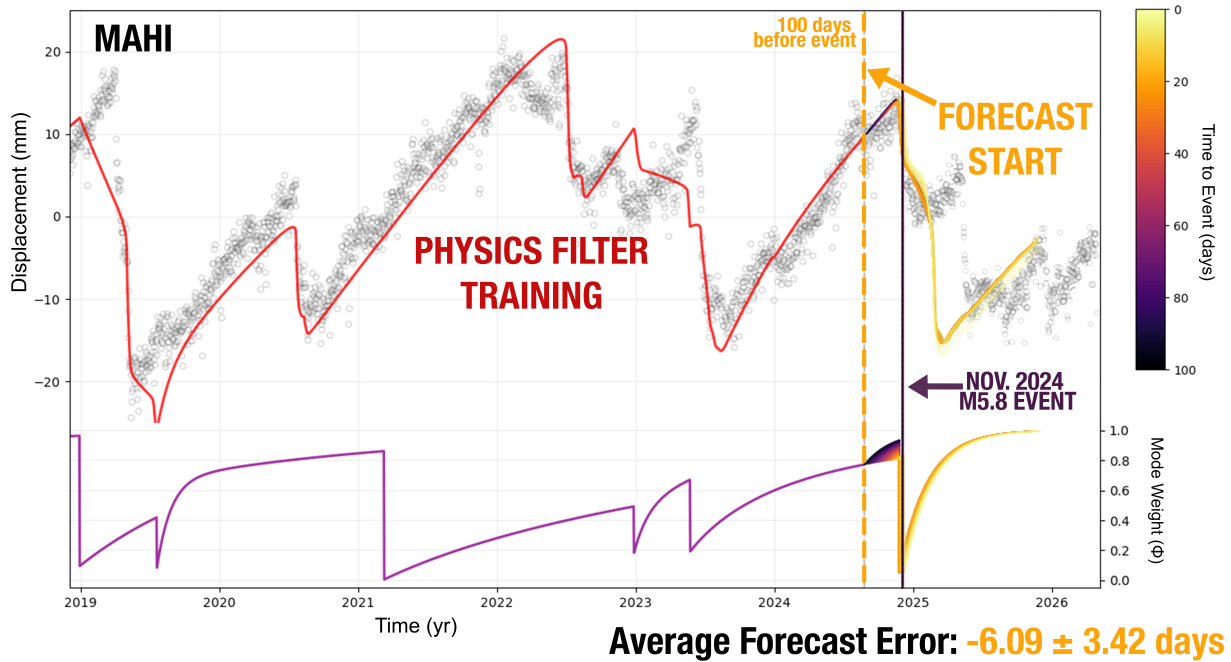} 
	\caption{\textbf{Deterministic phase space projection achieves high-accuracy timing predictions for the 2024 M5.8 slow slip event.}
    Detailed view of the MAHI station physics filter forecast over a 100-day initialization window (yellow dashed line), color-coded from 100 days out (dark blue) to the day of the event (light yellow).  (\textbf{Top}) Training data and displacement trajectories forecasted via k-nearest neighbors (KNN) phase space projection.  (\textbf{Bottom}) Corresponding evolution of the modal weight parameter ($\phi$).  The forecasting model dynamically calculates future $\phi$ states by monitoring the predicted displacement rate.  As the forecast projects natural tectonic loading, $\phi$ relaxes toward 1.  When the predicted displacement rate crosses a critical threshold, the model autonomously triggers a simulated rupture, resetting $\phi$ toward a baseline state near 0.  By strictly adhering to these deterministic mechanics, the phase space projection achieves an average event prediction error of $-6.09 \pm 3.42$ days.
    }
	\label{fig:forecastsphi} 
\end{figure}

Fig.~\ref{fig:forecastsphi} provides a magnified view of MAHI’s forecast immediately surrounding the event.  To fully confirm the autonomous predictive power of the model, we reimplemented the $\phi$ evolution variable.  Using the same ODE, we allow the system to continue evolving by tracking the predicted displacement rate rather than relying on discrete, pre-selected event dates as we did in training.  Once the predicted displacement rate drops below a critical threshold, the system autonomously signals a rupture, dropping $\phi$ back to 0.  Across 100 initialization windows, the predictions yielded a forecast accuracy of $-6.09 \pm 3.42$ days, essentially predicting the timing of the earthquake a week in advance.  

The reliability of these projections is fundamentally bounded by the Maximum Lyapunov Exponent ($\lambda_{max}$), which establishes the intrinsic limit of the resolved attractor. We find that the physics-based manifold effectively doubles the fundamental predictability horizon across the entire East Coast cluster, extending the regional deterministic ceiling to $37.9 \pm 4.3$ days (table~\ref{tab:final_mle}). Crucially, cumulative convergence analysis reveals that these horizons reach asymptotic stability across the decadal record, shifting by an average of only $1.2$ days over the final five years. This stability confirms that the predictability limit is a stationary, invariant mechanical property of the fault interface, rather than a statistical artifact or fluctuating signal-to-noise ratio.

These results provide compelling evidence that this multi-physical (thermo-poro-mechanical) framework captures the fundamental physics governing Earth's subduction mechanics.  While there is room for further refinement, such as localized parameterization, improved global optimization algorithms, and the implementation of a spatially coordinated network grid for triangulation and epicenter inversion, the core dynamics are successfully resolved. 

Ultimately, the convergence of these near-field forecasts across a decadal scale suggests that the perceived stochasticity of subduction transients is largely an artifact of instrumental noise obscuring a deterministic skeleton. By isolating this skeleton through a spatially sensitive, physics-based manifold, we demonstrate that subduction environments follow a predictable trajectory, moving geophysics towards a predictive horizon akin to atmospheric forecasting.


\begin{figure}
	\centering
	\includegraphics[width=1\textwidth]{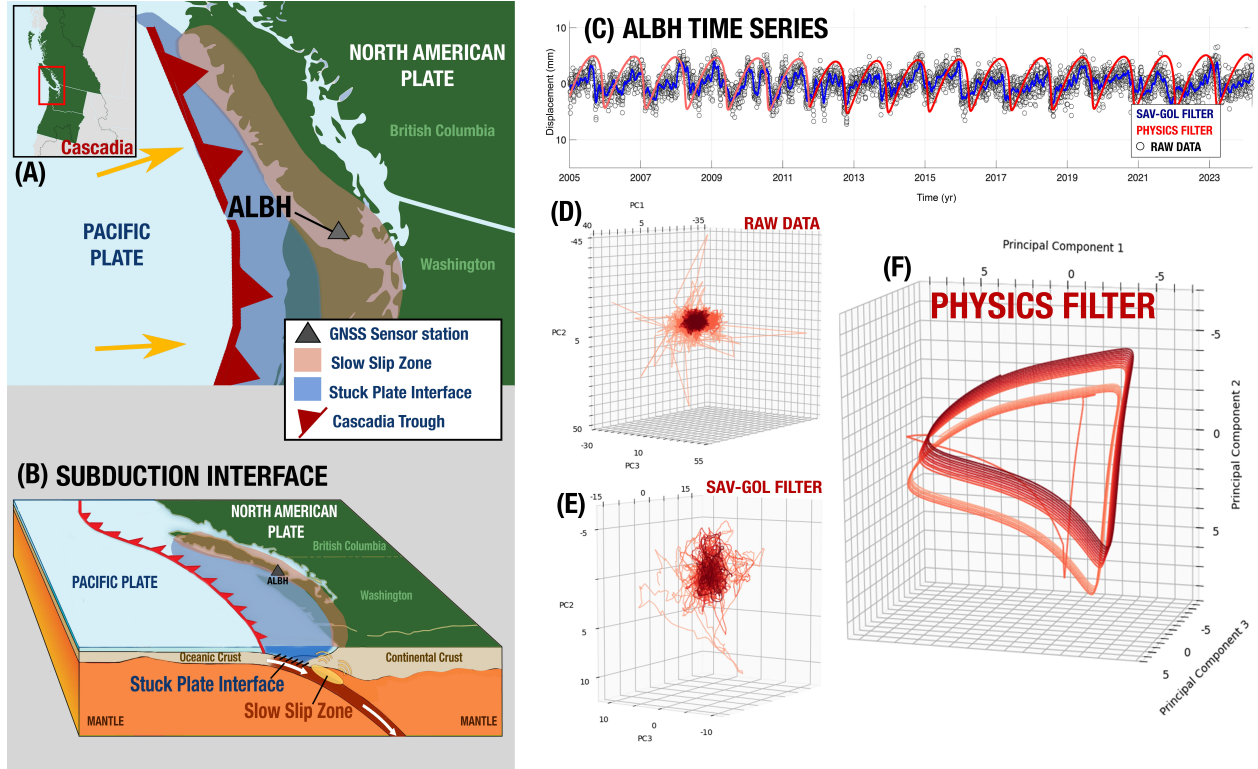} 
  \caption{\textbf{Physics-based filter isolates periodic slow slip dynamics in the Cascadia subduction zone.}
		(\textbf{A and B}) Map and 3D schematic of the Cascadia margin.  Slow Slip Events (SSEs), the fundamental drivers of instability in subduction zones, are consistently observable by the GNSS sensors located over the Slow Slip Zone (light orange). (\textbf{C}) ALBH displacement time series overlaid with the physics-based filter (red).  Due to the lack of major local earthquakes (magnitude $ > 5$), the physics filter, which models pressure and temperature within the subduction zone, perfectly aligns with the SSE signal.  The transition from light to dark red represents a non-physical sensor jump. (\textbf{D to F}) Phase space reconstructions from the raw displacement data, simple Savitzky-Golay filter, and the physics filter, respectively.  The physics filter extracts a remarkably cleaner attractor, demonstrating the underlying deterministic thermo-poro-mechanical dynamics.}
	\label{fig:CASC_map} 
\end{figure}

\begin{figure}
	\centering
	\includegraphics[width=0.6\textwidth]{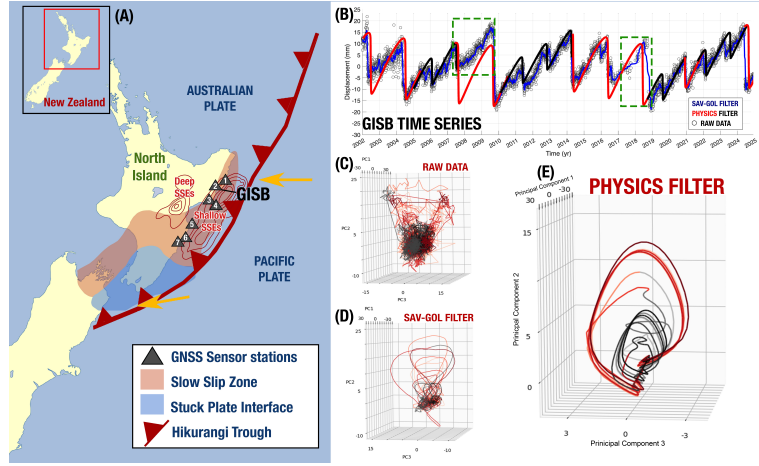} 
	\caption{\textbf{Physics-based filter isolates bimodal slow slip dynamics in Hikurangi margin.}
  (\textbf{A}) Map of the Hikurangi subduction zone, with the seven GNSS stations (1. CNST, 2. GISB, 3. KOKO, 4. MAHI, 5. CKID, 6. PAWA, 7. PORA) positioned over the active Slow Slip Zone (light orange). Deep and shallow SSE contours are shown, more details on their slip magnitudes and depth can be found in Fig. 2 of \cite{poulet2026}.  (\textbf{B}) GISB displacement time series overlaid with the physics filter (red and black).  Unlike Cascadia, Hikurangi's complex and seismically active margin oscillates between two distinct modes driven by competing deep and shallow SSEs.  These modes differ only through the boundary (regional) stresses that the overriding plate applies to the subduction interface.  The fundamental red mode (lower stresses) represents deeper, diffusion-dominant processes.  The system naturally loads into the black mode (higher stresses), reflecting shallower, heat-generation-dominated cycles.  External excitation (such as a local seismic event) is then needed to release the excess stress and return to the fundamental red mode.  Green dashed boxes represent a non-physical sensor jump and an area of data sparsity. (\textbf{C to E}) Phase space reconstructions from the raw displacement data, simple Savitzky-Golay filter, and the physics filter, respectively. The physics filter reveals a clean, low-dimensional bimodal attractor, confirming the presence of deterministic dynamics governing this subduction zone.}
	\label{fig:NZ_map} 
\end{figure}

\clearpage 

\bibliography{citations} 
\bibliographystyle{sciencemag}
%
%



\section*{Acknowledgments}

\paragraph*{Funding:}
This work was supported by the U.S. National Science Foundation (NSF Projects CMMI-2042325 and 2332069).

\paragraph*{Author contributions:}
M. Veveakis and T. Poulet created the multi-physical model.
V. Keane, under the mentorship of M. Veveakis, performed all data processing, automated event optimization, phase space reconstructions, machine learning forecasting implementations, and associated analyses.  V. Keane and M. Veveakis wrote the manuscript with input from T. Poulet.

\paragraph*{Competing interests:}
There are no competing interests to declare. 

\paragraph*{Data and materials availability:}
All geodetic and seismic data utilized in this study are publicly available from the GeoNet sensor network (https://www.geonet.org.nz/) and the USGS Federation of Digital Seismograph Network (FSDN). The complete, reproducible code base and pipeline are publicly archived on GitHub at https://github.com/vkeane29/slow-slip-forecasting.git. There are no restrictions on data availability or material transfer agreements.


\subsection*{Supplementary materials}
Materials and Methods\\
Supplementary Text\\
Figs. S1 to S7\\
Tables S1 to S8\\
References \textit{(7-\arabic{enumiv})}\\ 

\newpage


\renewcommand{\thefigure}{S\arabic{figure}}
\renewcommand{\thetable}{S\arabic{table}}
\renewcommand{\theequation}{S\arabic{equation}}
\renewcommand{\thepage}{S\arabic{page}}
\setcounter{figure}{0}
\setcounter{table}{0}
\setcounter{equation}{0}
\setcounter{page}{1} 


\begin{center}
\section*{Supplementary Materials for\\ \scititle}

V.~Keane$^{1\ast}$,
	T.~Poulet$^{2}$,
    E.~Veveakis$^{1}$\and
	\\
    \small$^{1}$Department of Civil and Environmental Engineering, Duke University, Durham NC, USA.\and
    \\
	\small$^{2}$ Earth Process Models, CSIRO, Kensington WA, Australia.\and
\end{center}

\subsubsection*{This PDF file includes:}
Materials and Methods\\
Supplementary Text\\
Figures S1 to S7\\
Tables S1 to S8

\newpage


\subsection*{Materials and Methods}

\subsubsection*{Data Acquisition, Preprocessing, and Noise Analysis}

To isolate transient tectonic signals from the background secular motion of the Hikurangi Subduction margin, we implemented a multi-stage filtering pipeline on GNSS position time series harvested from the GeoNet sensor network \cite{geonet}. We focused on the East Coast cluster (GISB, KOKO, MAHI, PAWA, CKID, CNST, and PORA) due to their proximity to the active SSE zone.  For all stations, we utilized only the eastern component of the displacement, as it aligns with the dominant trench-perpendicular slip vector of the subduction interface.

Raw displacement $u(t)$ was corrected for secular tectonic drift via first-order linear detrending, followed by a nonlinear least-squares regression to remove multi-harmonic seasonal loading--primarily hydrological and atmospheric (Supplementary Text) \cite{liu2020}. Because standard linear filters risk over-smoothing the abrupt nonlinear transitions characteristic of SSEs, we performed a spectral noise analysis to characterize the instrumental and environmental jitter. Using a Lomb-Scargle periodogram to account for the uneven sampling inherent in geodetic observations, we estimated the Power Spectral Density (PSD) and derived the spectral index ($\alpha$) for each station (Supplementary Text) \cite{lomb, scargle}.

As shown in Table~\ref{tab:noise_stats}, our cluster consistently exhibits a spectral index of $\alpha \approx 0.8\text{–}0.9$. This near-unity index confirms that the noise floor is dominated by temporally correlated flicker noise rather than white noise. In this regime, standard smoothing techniques are prone to overfitting geodetic noise, creating spurious artifacts that mimic tectonic transients \cite{Williams}. This finding validates our deterministic approach. By grounding our filter in a physics-based model, we ensure the processed signal is constrained to the low-dimensional manifold of the fault interface, effectively separating deterministic physics from stochastic correlated noise.

\subsubsection*{Automated Event Selection}

Determining when a fault "resets" its dynamical state is a non-trivial challenge in noisy geodetic data. To avoid the subjectivity of manual event selection, we developed an automated spatiotemporal pipeline that identifies seismic triggers while strictly penalizing over-parameterization (Supplementary Text). Potential triggers were initially harvested from the USGS Federation of Digital Seismograph Networks (FDSN) catalog for magnitudes greater than 5.0 within a 300km radius of the GNSS cluster \cite{usgs}.

We recognized that high-frequency seismic swarms could force the optimization into fitting short-term aftershock sequences rather than the long-term recovery of the fault. To prevent this, triggers within a 180-day window were grouped into single "dynamical epochs," ensuring sufficient temporal resolution for the characteristic recovery timescale ($\tau$) of the subduction interface.

To distinguish tectonic responses from geodetic noise, we performed a sensitivity analysis on a rolling-median saliency threshold (Supplementary Text). A 2.5 mm threshold provided the optimal balance between signal capture and model efficiency, exceeding the cluster's 1.0–1.5 mm daily precision floor \cite{Williams}. Additionally, any event $M \geq 7.0$ was automatically enforced as a regional reset, acknowledging that massive energy releases can perturb the entire subduction interface and trigger mode switches even if local displacement is subtle. Final event configurations (Table~\ref{tab:event_dates}) were validated using the Akaike Information Criterion (AIC) to ensure that each reset was statistically justified by a significant reduction in residual variance rather than increased model complexity (Supplementary Text) \cite{akaike1974}.

\subsubsection*{Modal Fitting Optimization}

To resolve the non-linear transitions between physical modes, we define the system state, $\phi(t)$, using a first-order relaxation ODE:
\begin{equation}
\dot \phi = \frac{1-\phi}{\tau}
\label{eq:relax_ode}
\end{equation}
where $\phi$ represents the modal weight order parameter and $\tau$ the characteristic recovery timescale. The total system response is a weighted combination of pre-computed physical kernels derived from integrated strain pulses: $u_{total} = \phi u_{black} + (1-\phi) u_{red}$ (Supplementary Text).

To ensure physical continuity, displacement $u_{\text{mod}}$ for each segment is calculated relative to the state of the previous epoch. We employed Differential Evolution \cite{storn}--a global optimization algorithm--to minimize a multi-penalty objective function $\mathcal{L}$:
\begin{equation}
\mathcal{L} = \sum_{i=1}^{n} (u_{\text{obs,i}} - u_{\text{mod,i}})^2 + 10 \cdot J + 5 \cdot PTP
\end{equation}
where $J$ and $PTP$ represent penalties for initial synchronization (jump) and peak-to-peak amplitude accuracy, respectively. All parameters were constrained to physically realistic intervals (Table~\ref{tab:param_bounds}). To quantify uncertainty, we executed an ensemble of 15 independent trajectories per station, utilizing the ensemble mean as the final deterministic filter (Fig.~\ref{fig:extrafits}A).

\subsubsection*{Phase Space Reconstruction}
The ensemble mean fits were used to reconstruct the system’s phase space through Taken’s Embedding Theorem \cite{takens1981}. We constructed delay vectors $\mathbf{Y}(t)$ to uncover the higher-dimensional manifold governing fault displacement:
\begin{equation}
\label{eq:delay_vectors}
\mathbf{Y}(t) = [y(t), y(t - \tau), y(t - 2\tau), \dots, y(t - (m - 1)\tau)]
\end{equation}
where $y(t)$ is the observed state at time $t$. The reconstruction parameters, time lag ($\tau_{\text{delay}}$) and embedding dimension ($m$), were optimized for each station.

We determined $\tau_{\text{delay}}$ using the first local minimum of the Mutual Average Information (MAI) function to ensure coordinate independence (Supplementary Text) \cite{fraser1986, abarbanel1993}. To identify the minimum dimension required to represent the phase space, we employed the False Nearest Neighbors (FNN) algorithm (Supplementary Text) \cite{kennel1992, abarbanel1993}. Although the algorithm formally identified $m=4$ for most stations, an artifact typical of highly correlated flicker noise ($\alpha \approx 0.9$), we adopted $m=3$ for the final manifold reconstruction. This choice is physically motivated by the alignment with the three governing state variables of our thermo-poro-mechanical framework--temperature, pore pressure, and stress--and is validated by the significant reduction in FNN percentage achieved by the physics-based filter (Table~\ref{tab:fnn_percentages}).

\begin{figure}
\centering
\includegraphics[width=0.6\textwidth]{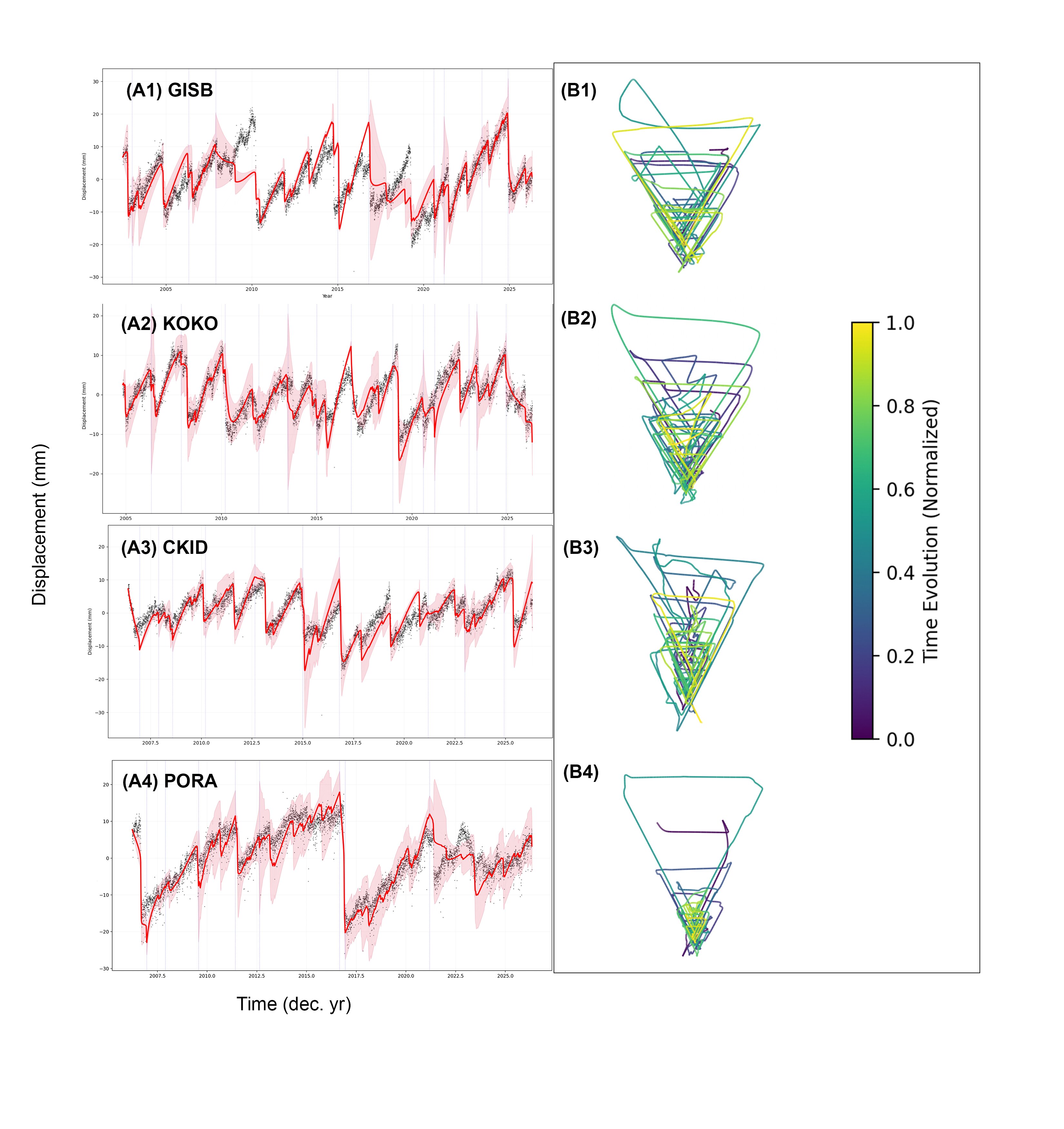}
  \caption{\textbf{Ensemble Optimization and Phase Space Reconstructions for additional sensors.}
  (\textbf{A}) Detrended GNSS displacement time series and corresponding physics-based ensemble fits for stations GISB, KOKO, CKID, and PORA. The red line represents the ensemble mean, while the shaded ribbon denotes the 95\% confidence interval. (\textbf{B}) Corresponding 3D phase space attractors following PCA rotation.}
\label{fig:extrafits}
\end{figure}

\begin{figure}
\centering
\includegraphics[width=0.6\textwidth]{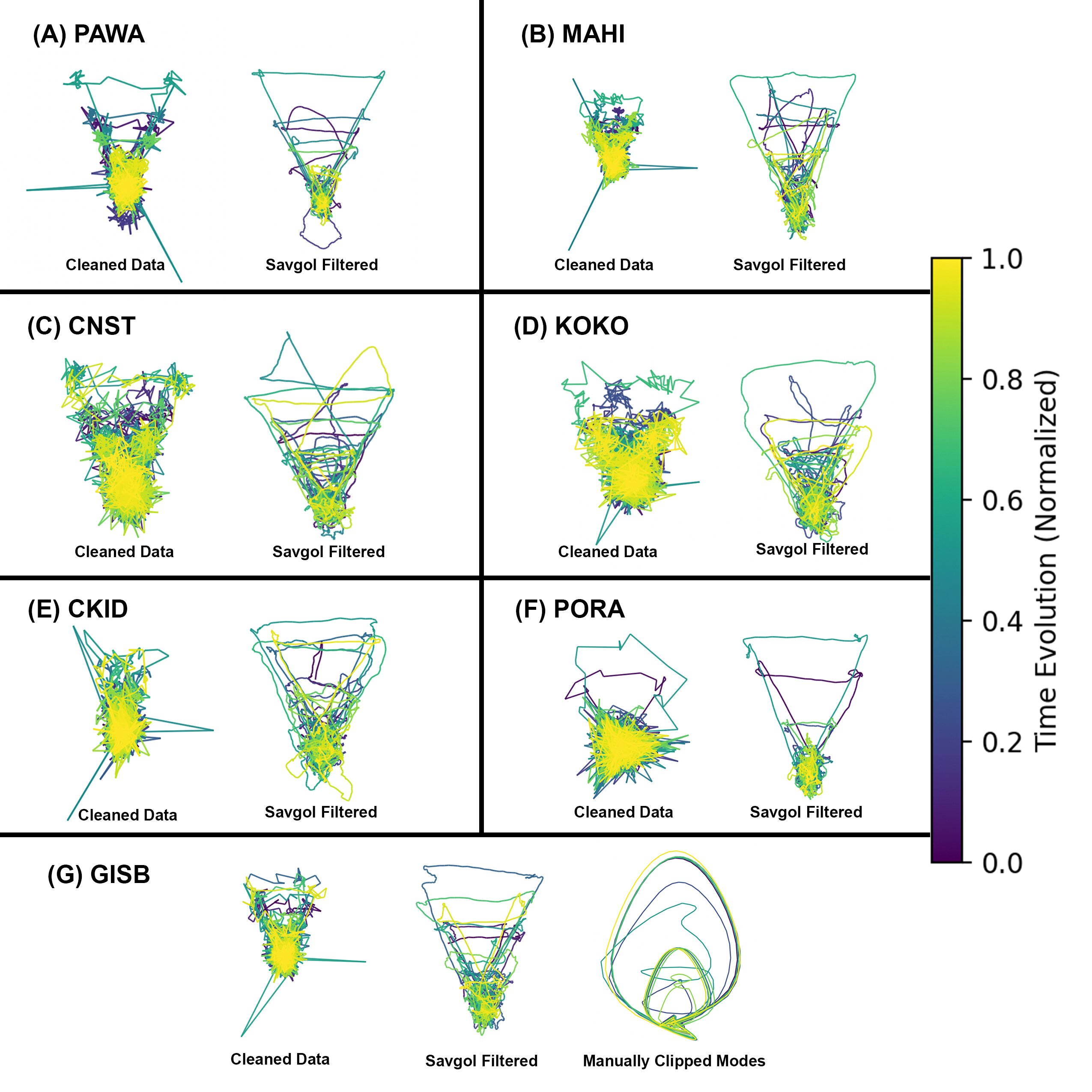}
\caption{\textbf{Comparison of Raw Residual and Standard Filter Phase Spaces.}
  Reconstructions of the attractors for all study stations using cleaned GNSS residuals and standard Savitzky-Golay filters. The manually clipped physical modes reconstruction for the GISB station is shown for comparison. All manifolds are subjected to the same PCA rotation to isolate nonlinear oscillations from secular tectonic drift.}
\label{fig:datafilter}
\end{figure}

To isolate nonlinear oscillations from secular tectonic drift, we applied a Principal Component Analysis (PCA) rotation to the embedded manifolds (Supplementary Text) \cite{abarbanel1993}. This transformation aligns the system along its principal axes, separating the high-variance background trend (PC 1) from the planar SSE cycles (PC 2 and 3), as visualized in Fig.~\ref{fig:CASC_map}F, Fig.~\ref{fig:NZ_map}F, Fig.~\ref{fig:forecasts}D, Fig.~\ref{fig:datafilter}, and Fig.~\ref{fig:extrafits}B.

\subsubsection*{Forecasting Implementation}

We project the current state, $\mathbf{Y}(t)$, along the deterministic manifold using a station-adaptive local linear K-Nearest Neighbors (KNN) approach \cite{fan2021}. For a given state, $\mathbf{Y}_{\text{now}}$, we identify $k$ historical neighbors, $\mathbf{Y}_j$, and their subsequent evolutions, $\mathbf{Y}_{j, \text{next}}$. We then solve for a local transition matrix $\mathbf{M}$, representing the linearized flow of the system, by minimizing the weighted least-squares error:

\begin{equation}
\min_{\mathbf{M}} \left( \sum_{j=1}^{k} w_j \left\| \mathbf{Y}_{j, \text{next}} - \mathbf{M} \begin{bmatrix} \mathbf{Y}_j \\ 1 \end{bmatrix} \right\|^2 + \alpha \|\mathbf{M}\|_F^2 \right)
\end{equation}

The forecast $\hat{\mathbf{Y}}(t+T)$ is generated by applying this optimal operator. The state vectors are augmented with a constant $1$ to allow $\mathbf{M}$ to capture affine translations, and $\|\mathbf{M}\|_F$ denotes the Frobenius norm. The weights $w_j$ are assigned via a tricube kernel based on the Euclidean distance in the reconstructed phase space, ensuring that closer analogs dictate the local geometry of the projection, while the $L_2$ regularization penalty ($\alpha$) prevents overfitting to local geodetic noise (Supplementary Text).

To ensure the model is predictive rather than descriptive, we utilize a walk-forward cross-validation strategy to optimize the neighbor count $k$ for each station independently. 

The intrinsic limits of predictability are defined by the Maximum Lyapunov Exponent ($\lambda_{max}$), calculated from the exponential divergence of nearest neighbor trajectories along the manifold \cite{abarbanel1993, sandri1996, wolf1985, rosenstein1993} (Supplementary Text).

To assess the longitudinal stability of the framework, we subjected the model to an ensemble test across five distinct tectonic epochs (2016.6627, 2016.9458, 2018.00, 2021.1886, and 2024.9233). We applied a temporal jitter of $\pm 2$ days to each forecast initialization date to calculate the predicted event error spread ($\sigma$), which serves as a robust proxy for the dynamical stability of the resolved manifold (Table~\ref{tab:stability_sigmas}).


\subsection*{Supplementary Text}

\subsubsection*{Full Nonlinear Coupled PDE System and Parameter Analysis}

The evolution of the fault interface is modeled as a 1D shear zone undergoing continuous tectonic loading. The physical mechanism governing the system is the thermally activated dehydration of serpentine minerals, a dominant driver of instability in subduction environments \cite{dobson, part3, wallace2004}. We adopt a thermo-poro-mechanical framework \cite{part1, part2, part3} defined by a coupled system of partial differential equations representing the evolution of normalized pore pressure ($\Delta p$) and temperature ($T$):

\begin{align}
\frac{\partial \Delta p}{\partial t} = \frac{\partial}{\partial z} \left[ \frac{1}{Le} \frac{\partial \Delta p}{\partial z} \right] + \frac{\Lambda}{M \sigma'_n} \frac{\partial T}{\partial t} + (1 - \phi_p)(1 - s) \zeta \mu_r e^{\frac{Ar \delta T}{1 + \delta T}} \\
\frac{\partial T}{\partial t} = \frac{\partial^2 T}{\partial z^2} + \left[ Gr (1 - \Delta p)^{-1/N} e^{\frac{a Ar}{1 + \delta T}} - (1 - \phi_p)(1 - s) \right] e^{\frac{Ar \delta T}{1 + \delta T}}
\label{eq:full_system}
\end{align}

The mathematical details as well as variable definitions can all be found in \cite{part1,part2,part3}.

The episodic nature of slow slip emerges directly from the competition between the diffusive and generative processes. As the overriding plate creeps, frictional deformation causes the temperature within the shear zone to rise. Because the parameter controlling frictional heating ($\alpha$) is strictly less than 1, mechanically driven heating dominates the early stages of the cycle before chemical processes initiate \cite{part1, part2}. Once the temperature breaches a critical Arrhenius activation threshold, a rapid, endothermic dehydration reaction is triggered. This reaction releases fluid from the solid rock skeleton, leading to a sharp spike in pore pressure ($\Delta p$), decreasing the effective stress across the interface, essentially lubricating the fault. This reduction in frictional resistance suppresses further deformational heating, preventing the interface from accelerating into a runaway seismic rupture \cite{part1, part2}.

The precise character of this limit cycle is dictated by $Gr$, which represents the ratio of heat generation relative to thermal diffusion and can be thought of as the boundary (regional) stresses that the overriding plate applies to the subduction interface.  In deeper, fundamental tectonic regimes, lower $Gr$ values yield a diffusion-dominated state. In these regimes, the broader variations in temperature and more dispersed fluid flow lead to larger displacement values and longer, multi-year recurrence periods. Conversely, as local stress gradients steepen--often in shallower zones--the system transitions into a generation-dominated regime characterized by higher $Gr$ values. Here, heat generation rapidly outpaces diffusion, shrinking the timescale of the limit cycle and producing the localized, high-frequency strain pulses typical of shallow SSEs \cite{part1, part2, part3}.

\subsubsection*{Strain File Derivation}

The underlying modes are derived from strain rates $\dot{\epsilon}$ stored in matrices pre-computed via a numerical PDE solver of the governing system (Eq.~\ref{eq:full_system}). The 1D thermo-poro-mechanical system is solved utilizing a rational spectral collocation method \cite{tee2006}. The spatial domain across the shear zone is discretized using 17 adaptively transformed Chebyshev grid points ($N=16$) to accurately capture the steep gradients generated by localized frictional heating. Exploiting fault symmetry, the system is solved on the half-space $[0, H]$ with Neumann boundary conditions enforced at the fault center. Time integration is performed using a stiff ordinary differential equation solver with a relative tolerance of $10^{-4}$ to accommodate the highly nonlinear reaction kinetics.

These numerical solutions yield distinct limit cycles depending on the prescribed Gruntfest number ($Gr$) \cite{part1, part2}. The fundamental, diffusion-dominated state (the red mode, lower $Gr$) exhibits higher displacement and strain amplitudes over a longer $\sim$2-year recurrence period, accompanied by broad temperature oscillations ranging from $270^\circ$C to $600^\circ$C. In contrast, the generation-dominated state (the black mode, higher $Gr$) is characterized by localized, lower-amplitude strain pulses on a faster 14-month cycle, with tighter temperature fluctuations between $330^\circ$C and $500^\circ$C.

The resulting discrete strain rate arrays, $\dot{\epsilon}(z, t')$, are then spatially and temporally integrated to produce the continuous displacement kernels:
\begin{equation}
u(t) = \int_{0}^{t} \left( \int_{-H}^{H} \dot{\epsilon}(z, t') dz \right) dt'
\end{equation}
where $H$ is the half-width of the shear zone. These resulting physical pulses are subsequently interpolated to match the daily sampling rate of the GNSS sensor network for the modal optimization pipeline.

\subsubsection*{Data Detrending and Noise Analysis}

To account for periodic environmental loading \cite{liu2020}, we utilize a nonlinear least-squares regression to model the seasonal components:
\begin{equation}
u_{seasonal}(t) = \sum_{k=1}^{2} a_k \sin(2\pi f_k t + p_k)
\end{equation}
where $f_1 = 1.0 \, \text{yr}^{-1}$ (annual) and $f_2 = 2.0 \, \text{yr}^{-1}$ (semi-annual). The amplitudes $a_k$ and phases $p_k$ are optimized per station. The long-term secular velocity, $u_{trend}$, is modeled as a linear function of time:
\begin{equation}
u_{trend}(t) = v t + b
\end{equation}
where $v$ is the constant inter-seismic velocity and $b$ is the intercept. The cleaned residuals, $u_{clean} = u_{raw} - u_{trend} - u_{seasonal}$, are then subjected to spectral analysis.

The spectral index, $\alpha$, is derived from the Power Spectral Density (PSD) estimated via a Lomb-Scargle periodogram, which is robust against the unevenly sampled gaps common in daily geodetic observations \cite{lomb, scargle}. We fit a power law in log-log space:
\begin{equation}
\ln(PSD(f)) = -\alpha \ln(f) + C
\end{equation}
where $f$ is the frequency. The resulting $\alpha$ values ($0.78 \leq \alpha \leq 0.96$) confirm a persistent non-white noise regime (Table~\ref{tab:noise_stats}). Consequently, any observed transient must be reconciled with our deterministic physical framework to verify its tectonic origin.

\subsubsection*{Event Filtering and Statistical Validation}

The local GNSS sensor saliency is defined as the absolute difference between the median displacement $u$ in a 14-day pre-event window and a 14-day post-event window:

\begin{equation}
  \Delta u_{saliency} = | \text{median}(u_{post}) - \text{median}(u_{pre}) |
\end{equation}

By using a 14-day median rather than single-day offsets, we mitigate the impact of transient outliers and common-mode errors \cite{Williams}. Sensitivity analysis revealed that a 2.5 mm threshold maximized $R^2$ for top-performing stations by filtering sub-threshold flicker noise that otherwise interferes with the recovery ODE optimization.

The Akaike Information Criterion (AIC) was employed to resolve the trade-off between model accuracy and complexity \cite{akaike1974}. For each GNSS station, the AIC is calculated as:
\begin{equation}
AIC = 2K + n \ln\left(\frac{SSR}{n}\right)
\end{equation}
where $SSR$ is the sum of squared residuals and $n$ is the number of daily observations. The total degrees of freedom, $K$, for a station model is defined by the number of dynamical segments ($N_{seg}$) and discrete resets ($N_{res}$):
\begin{equation}
K = (N_{seg} \times 7) + N_{res}
\end{equation}
Here, 7 represents the internal parameters per segment (timescale $\tau$, modal shifts/scaling, and initial state $\phi_{init}$). This penalized likelihood approach ensures that the introduction of a new dynamical segment is statistically justified.

\subsubsection*{Modal Optimization Accuracy and Uncertainty Quantification}

To evaluate the reliability of the deterministic fits, we utilize a dual-metric approach. Global accuracy is quantified via the Coefficient of Determination ($R^2$) and Root Mean Square Error (RMSE). To account for the sensitivity of the nonlinear optimization to noise, we define the "model disagreement" ($\sigma_{\text{mod}}$) as the standard deviation across the 15-run ensemble. A low $\sigma_{\text{mod}}$ acts as a proxy for the stability of the physical manifold, indicating that the governing PDEs converge on a singular dynamical path despite the noisy background.

Total uncertainty ($\sigma_{\text{total}}$) is calculated as the quadrature sum of the model disagreement and the station-specific noise floor ($\sigma_{\text{noise}}$):
\begin{equation}
\sigma_{\text{total}} = \sqrt{\sigma_{\text{mod}}^2 + \sigma_{\text{noise}}^2}
\end{equation}
This provides conservative error bounds for subsequent forecasting, ensuring projections are grounded in high-confidence regions of the attractor.

\subsubsection*{Mutual Average Information (MAI)}

To select a time delay $\tau$ that avoids over-correlation between coordinates, we calculate the Mutual Average Information (MAI):

\begin{equation}I(\tau) = \sum_{i,j} P(u(t), u(t+\tau)) \ln \frac{P(u(t), u(t+\tau))}{P(u(t))P(u(t+\tau))}
\end{equation}

where $P(u(t))$ and $P(u(t+\tau))$ are the marginal probability distributions of the displacement signal, and $P(u(t), u(t+\tau))$ is the joint probability density. The first local minimum defines the lag where the delayed coordinates are maximally independent while remaining part of the same dynamical trajectory \cite{fraser1986, abarbanel1993}.

\subsubsection*{False Nearest Neighbors (FNN)}

We use the False Nearest Neighbors (FNN) method to determine the minimum dimension $m$ required to unfold the attractor without self-intersection \cite{kennel1992, abarbanel1993}. The criterion for a false neighbor is defined as:
\begin{equation}
\frac{|u(t+m\tau) - u^{(NN)}(t+m\tau)|}{||\mathbf{Y}_m(t) - \mathbf{Y}_m^{(NN)}(t)||} > R_{tol}
\end{equation}
where the numerator represents the distance between a point and its nearest neighbor when projected into the $(m+1)$-th dimension and $R_{tol} = 0.1$ or 10\%. Highly correlated flicker noise ($\alpha \approx 0.9$) typically inflates FNN counts by creating artificial self-intersections. Constraining the system to $m=3$ using the physics-based filter reduces the average FNN percentage from 28.12\% (raw) to 13.94\%, confirming that the additional dimensionality derived is a manifestation of instrumental noise rather than fault physics.

\subsubsection*{Principal Component Analysis (PCA) Rotation}

To orient the manifold for analysis, we construct a trajectory matrix $\mathbf{X}$ from the reconstructed state vectors:

\begin{equation}
\mathbf{X} = 
\begin{bmatrix} 
\mathbf{Y}(t_1) \\ \mathbf{Y}(t_2) \\ \vdots \\ \mathbf{Y}(t_n) \end{bmatrix} 
\end{equation}

We perform a Singular Value Decomposition (SVD) on this mean-centered matrix to identify the directions of maximum variance \cite{broomhead1986, abarbanel1993}. We then apply the orthogonal transformation $\mathbf{Y}_{rot} = \mathbf{X} \mathbf{V}$, where $\mathbf{V}$ is a rotation matrix whose columns define an orthonormal basis. In this rotated frame, PC1 isolates the secular tectonic drift, while PC2 and PC3 reveal the underlying SSE cycles as distinct, planar oscillations.

\subsubsection*{Local Linear KNN and Regularization}

The deterministic forecasting utilizes a local linear approximation of the phase space flow \cite{fan2021}. To transition from descriptive fitting to true walk-forward forecasting, the optimal neighbor count ($k \in [60, 200]$) is determined dynamically per station using a 10-day historical validation window. To prevent singular matrices and mitigate the influence of distant neighbors, we employ Ridge regression with a regularization parameter of $\alpha = 0.1$. The local geometry is enforced via a tricube distance weighting function:
\begin{equation}
w_j = \left(1 - \left(\frac{d_j}{d_{max}}\right)^3\right)^3
\end{equation}
where $d_j$ is the Euclidean distance to the $j$-th neighbor, and $d_{max}$ is strictly defined as $1.01 \times \max(d_j)$ to prevent division by zero for identical states.

\subsubsection*{Predictability Horizon and the Theiler Window}
The reliability of these projections is fundamentally bounded by the Maximum Lyapunov Exponent ($\lambda_{max}$), which establishes the rate of exponential divergence of trajectories within the attractor:
\begin{equation}
d(t) \approx d_0 e^{\lambda_{max} t} 
\end{equation}

To estimate $\lambda_{max}$ directly, we employ the Rosenstein algorithm \cite{rosenstein1993}. We compute the average logarithmic rate of divergence between nearest neighbors across the reconstructed manifold:
\begin{equation}
\lambda_{max} = \frac{1}{N \Delta t} \sum_{i=1}^{N} \ln \left( \frac{d_i(t + \Delta t)}{d_i(t)} \right)
\end{equation}
where $N$ is the total number of reference states evaluated, $d_i(t)$ is the initial Euclidean distance between the $i$-th reference state and its nearest neighbor, and $d_i(t + \Delta t)$ is the distance between their respective trajectories after a forward evolution time $\Delta t$.

Crucially, to ensure this calculation reflects the deterministic physical structure of the subduction interface rather than short-term temporal autocorrelation, we enforced a Theiler window \cite{theiler1986} of $W_T = m \times \tau$ during the nearest-neighbor selection. This restriction forbids the algorithm from selecting temporally adjacent points as dynamical analogs, guaranteeing that neighbors represent distinct tectonic cycles.

We assessed the stability of $\lambda_{max}$ through an expanding-window convergence test. By calculating $T_h \approx 1/\lambda_{max}$ across five cumulative observational windows, we observe that the physics-based horizons reach asymptotic stability as the record length increases, typically shifting by $1.2$ days between the 2021 and 2024 epochs. The reported horizons in Table~\ref{tab:final_mle} represent the mean $\pm$ one standard deviation across these tectonic regimes. This stability indicates that $T_h$ is a stationary geophysical property of the fault patches—governed by the invariant thermo-poro-mechanical state of the interface--whereas raw data horizons remain trapped within a significantly lower, noise-dominated regime ($13.1 \pm 1.0$ days).

\subsubsection*{Objective Event Detection}

To eliminate subjective bias when evaluating forecast accuracy against observed tectonic epochs, we implemented an objective, gradient-based event detection algorithm. The onset of a predicted slip event is defined as the moment the phase space velocity (the first derivative of the predicted displacement, $\frac{dy}{dt}$) drops below a critical threshold relative to the inter-seismic baseline. This threshold was universally defined as two standard deviations below the mean velocity of the quiescent period ($\mu - 2\sigma$). If a forecast trajectory fails to breach this threshold, it is classified as a non-nucleation state, directly supporting the model's capacity for correct rejection during the 2018 control epoch.

\subsubsection*{Longitudinal Stability and Temporal Jittering}

Standard accuracy metrics (e.g., RMSE, MAE) are highly sensitive to the exact initialization date of a forecast. To rigorously test the physical discipline of the manifold, we introduced a temporal jitter to the initialization vector. For each tectonic epoch evaluated in the stability analysis, the forecast start date was shifted across a $\pm 2$ day window. The standard deviation of the resulting predicted event dates yields the stability $\sigma$. A low $\sigma$ indicates that the system is tightly constrained by the thermo-poro-mechanical laws of the PDE system, naturally guiding adjacent initial conditions back onto a singular, deterministic trajectory. 

When subjected to this temporal jittering, forecasts relying on raw or smoothed data deteriorated rapidly during non-event windows, drifting toward extremely high instability values of $\sigma \sim 26-37$ days for the primary stations (Table~\ref{tab:stability_sigmas}). Conversely, our physics-based model correctly identified the absence of a physical bifurcation across the network, signaling a stable non-nucleation state. This ensures that the model only projects a slip trajectory when the fundamental state variables breach the critical thresholds defined by the governing PDEs.

\subsubsection*{Software and Data Availability}

The complete code base is publicly available on GitHub at \url{https://github.com/vkeane29/slow-slip-forecasting.git}. The computational pipeline is implemented in Python, utilizing standard scientific libraries including Pandas, NumPy, SciPy, and Scikit-learn for data manipulation, optimization, and modeling. Time-series and nonlinear dynamics analyses are handled via Statsmodels and NoLiTSA, while visualization relies on Matplotlib and mpl\_toolkits. 

The repository is structured sequentially to reflect the analytical stages of the study:
\begin{itemize}
    \item \textbf{\texttt{/data\_downloading}:} \texttt{load\_data.py} and \texttt{get\_timeser.py} handle the automated retrieval of GeoNet GNSS data based on specified station codes, isolating and saving individual station time series.
    \item \textbf{\texttt{/preprocess\_data}:} \texttt{clean\_data.py} evaluates station-specific noise profiles, performs secular and seasonal detrending, and outputs the cleaned residuals. \texttt{event\_select.py} then applies the automated saliency filter to identify discrete seismic triggers and define dynamical epochs.
    \item \textbf{\texttt{/phys\_model}:} \texttt{model\_clipping.py} integrates the pre-computed PDE \texttt{strain\_files}. It executes the 15-run Differential Evolution ensemble optimization and extracts the final averaged deterministic manifold fits.
    \item \textbf{\texttt{/phase\_space\_recon}:} \texttt{phase\_space\_params.py} calculates the optimal time delay ($\tau$) and embedding dimension ($m$) across the physics-based model, Savitzky-Golay filter, and raw data. \texttt{all\_phase\_spaces.py} subsequently constructs the 3D phase space attractors.
    \item \textbf{\texttt{/forecasting}:} This module contains the core predictive algorithms and stability metrics:
    \begin{itemize}
        \item \texttt{rolling\_forecast.py}: Generates local linear forecasts for specified stations and target events across the three data processing methods, producing comparative figures over specific initialization windows (e.g., 30 days and 17.5 days prior to an event).
        \item \texttt{stability\_forecasts.py}: Evaluates the longitudinal stability of the manifold across multiple tectonic epochs and stations. It calculates forecast convergence and rigorously tests the model's capacity for correct rejection (avoiding false positives during non-nucleation states).
        \item \texttt{MLE\_analysis.py}: Calculates the Maximum Lyapunov Exponent ($\lambda_{max}$) to establish deterministic predictability horizons, iterating over multiple years of the record to verify asymptotic stability.
        \item \texttt{phi\_forecast.py}: Calculates the autonomous timing predictions by driving the reduced-order $\phi$ evolution ODE with predicted displacement rates, allowing the system to naturally trigger discrete resets without pre-selected event dates.
    \end{itemize}
\end{itemize}


\clearpage

\begin{figure}
\centering
\includegraphics[width=0.6\textwidth]{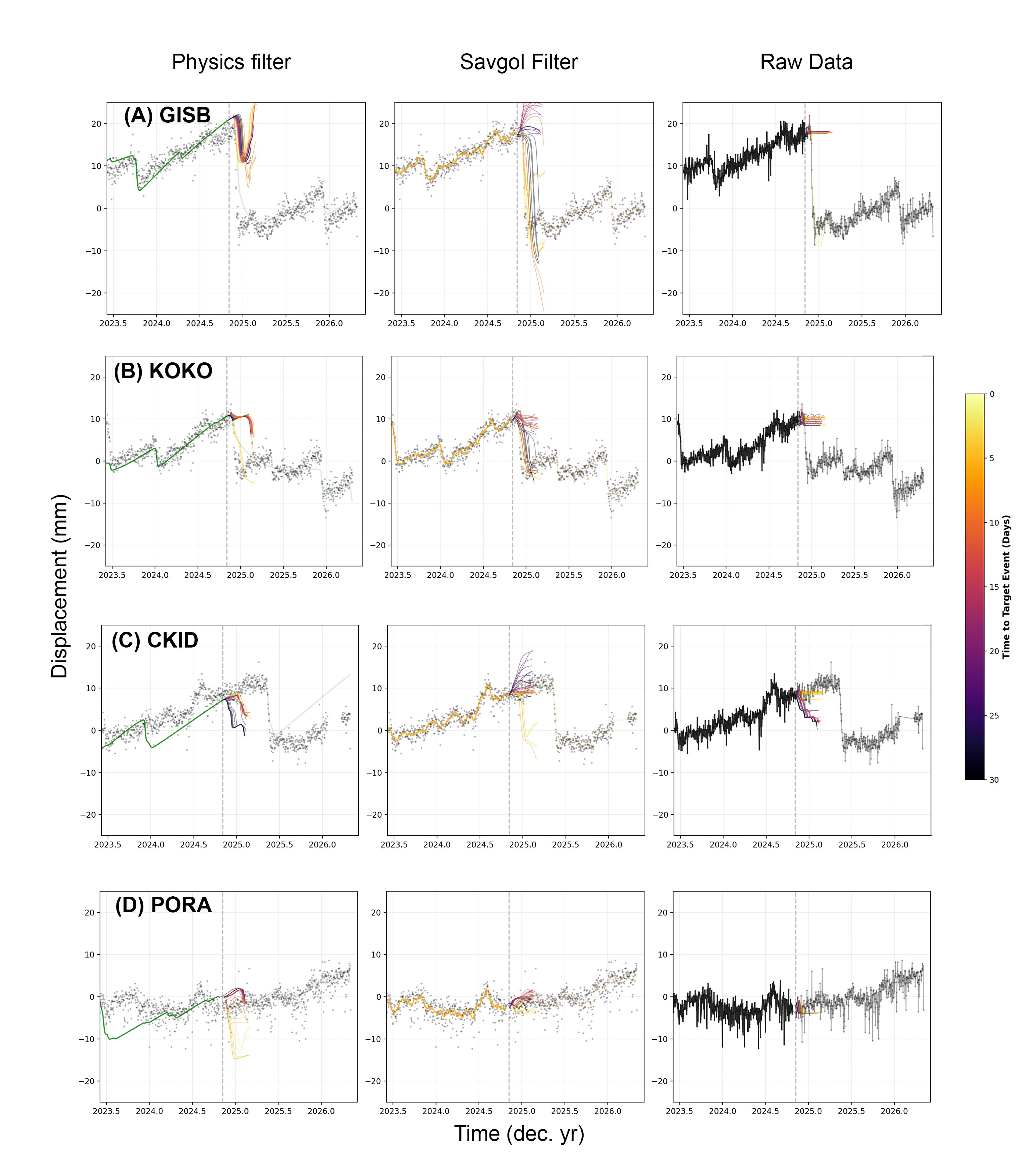}
\caption{\textbf{Additional Station Forecasts for 2024 $M5.8$ event.} Rolling 30-day forecasts for the additional East Coast stations (GISB, KOKO, CKID, PORA) targeting the 2024 $M5.8$ northern event. The physics-based filter, Savitzky-Golay filter, and cleaned residual forecasts are shown. Vertical dashed lines indicate the forecast initialization dates relative to the event onset. Forecasts are extended 90 days past the forecast start date.}
\label{fig:extra_2024}
\end{figure}

\begin{figure}
\centering
\includegraphics[width=0.6\textwidth]{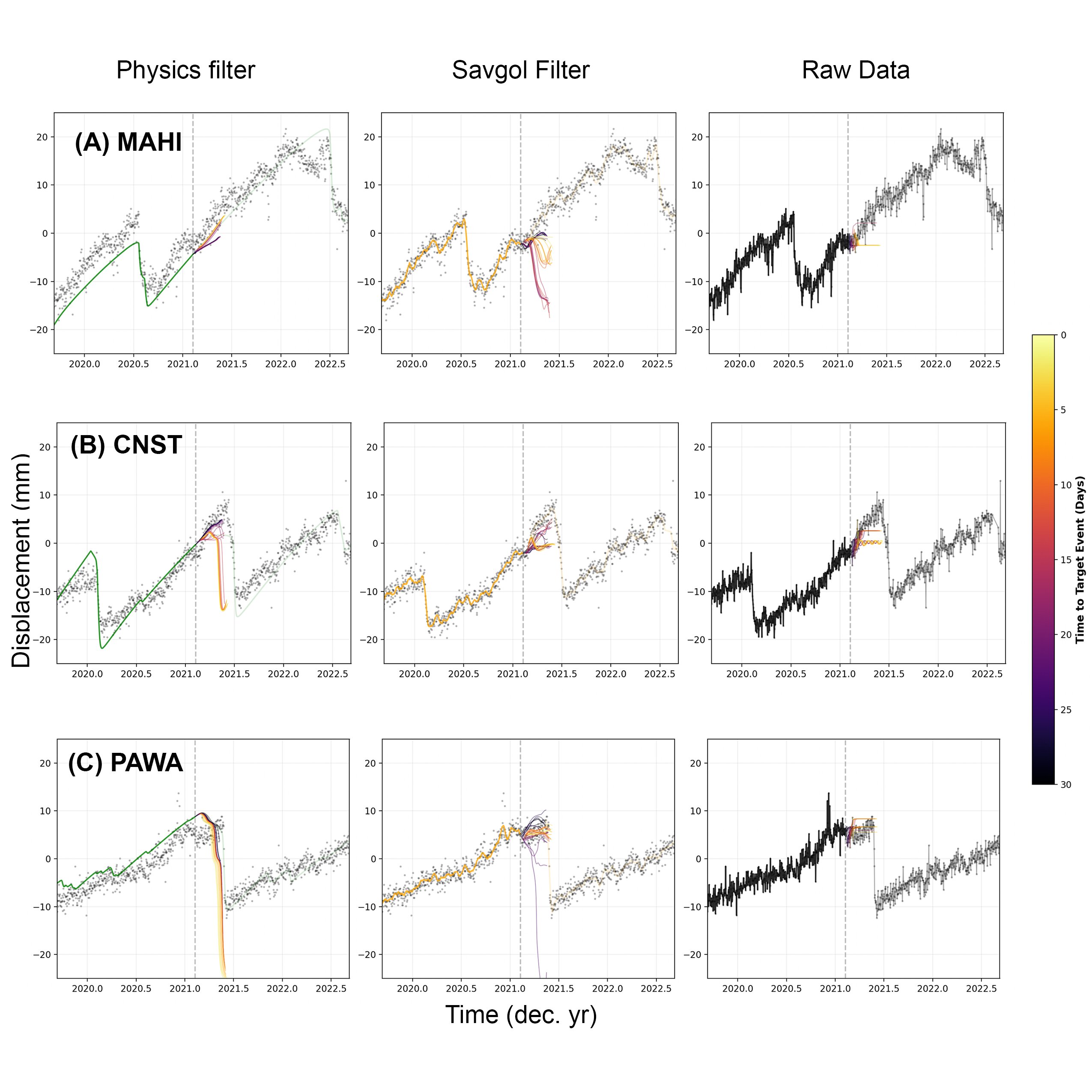}
\caption{\textbf{Top Performing Station Forecasts for 2021 $M7.3$ event.} Rolling 30-day forecasts for the top-performing East
Coast stations (MAHI, CNST, PAWA) targeting the 2021 $M7.3$ northern event. The physics-based filter, Savitzky-Golay filter, and cleaned residual forecasts are shown. Vertical dashed lines indicate the forecast initialization dates relative to the event onset. Forecasts are extended 90 days past the forecast start date.}
\label{fig:best2021}
\end{figure}

\begin{figure}
\centering
\includegraphics[width=0.6\textwidth]{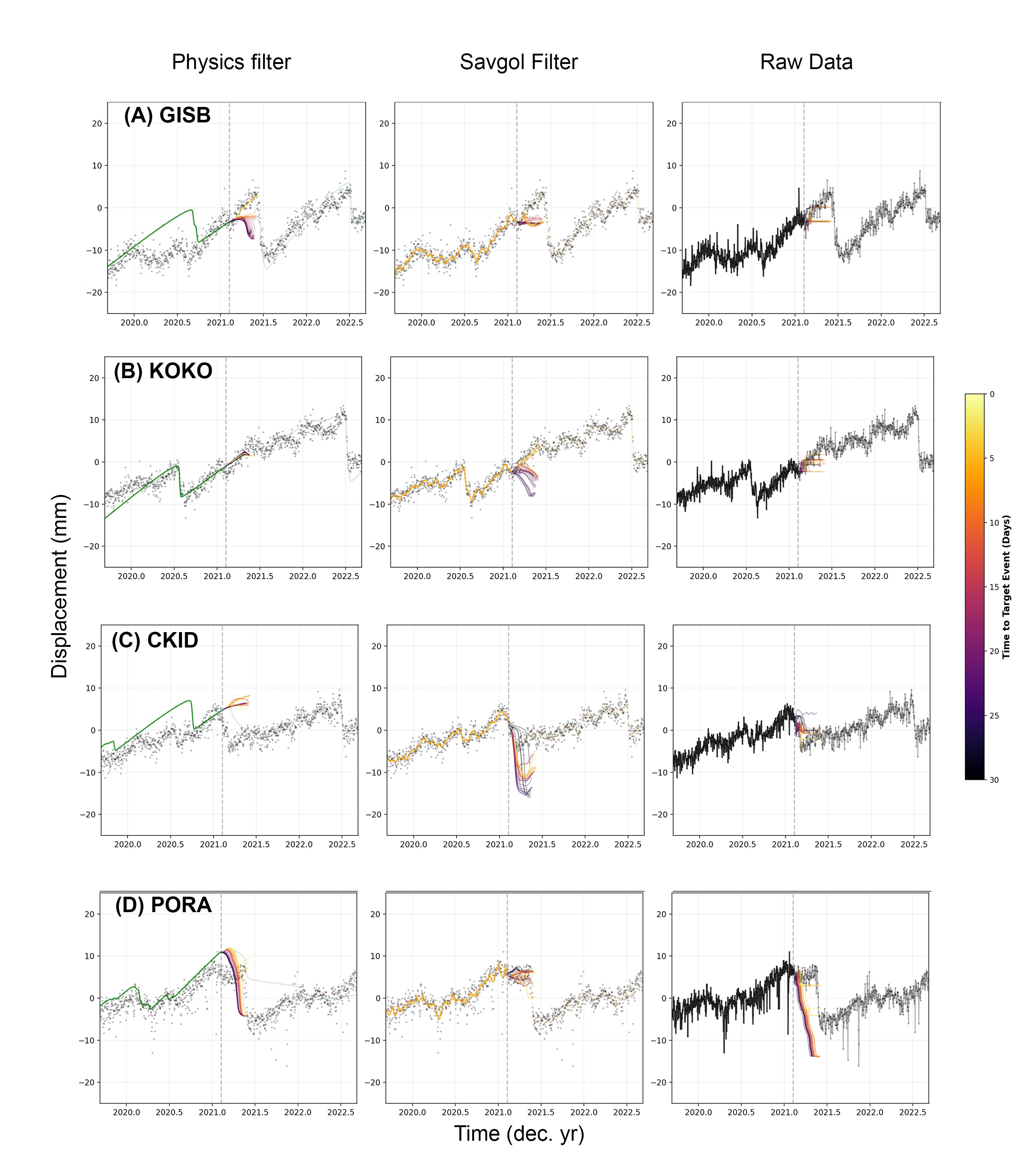}
\caption{\textbf{Additional Station Forecasts for 2021 $M7.3$ event.} Rolling 30-day forecasts for the additional East
Coast stations (GISB, KOKO, CKID, PORA) targeting the 2021 $M7.3$ northern event. The physics-based filter, Savitzky-Golay filter, and cleaned residual forecasts are shown. Vertical dashed lines indicate the forecast initialization dates relative to the event onset. Forecasts are extended 90 days past the forecast start date.}
\label{fig:extra2021}
\end{figure}

\begin{figure}
\centering
\includegraphics[width=0.6\textwidth]{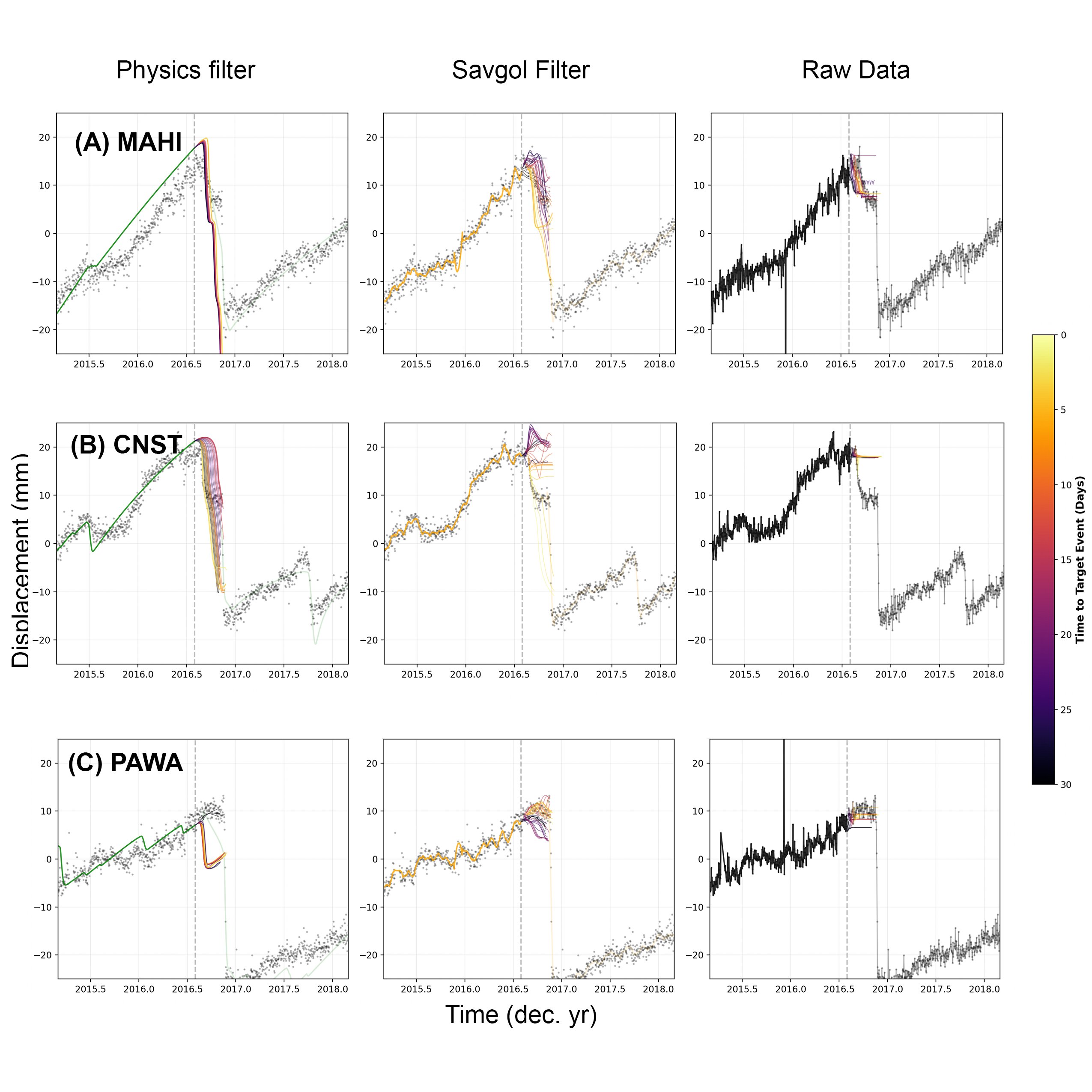}
\caption{\textbf{Top Performing Station Forecasts for 2016 $M7.0$ event.} Rolling 30-day forecasts for the top-performing East
Coast stations (MAHI, CNST, PAWA) targeting the 2016 $M7.0$ northern event. The physics-based filter, Savitzky-Golay filter, and cleaned residual forecasts are shown.Vertical dashed lines indicate the forecast initialization dates relative to the event onset. Forecasts are extended 90 days past the forecast start date.}
\label{fig:best2016}
\end{figure}

\begin{figure}
\centering
\includegraphics[width=0.6\textwidth]{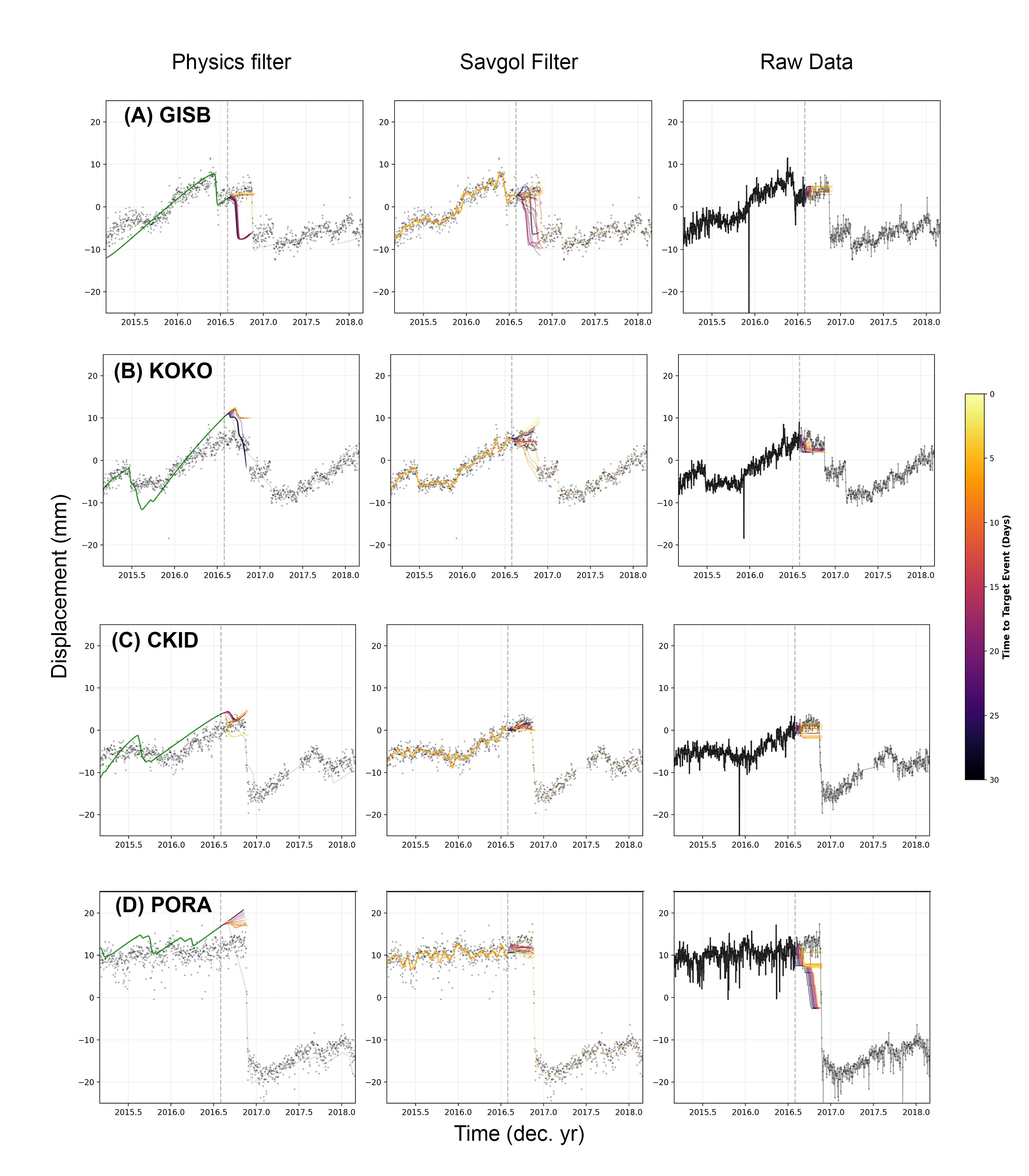}
\caption{\textbf{Additional Station Forecasts for 2016 $M7.0$ event.} Rolling 30-day forecasts for the additional East
Coast stations (GISB, KOKO, CKID, PORA) targeting the 2016 $M7.0$ northern event. The physics-based filter, Savitzky-Golay filter, and cleaned residual forecasts are shown. Vertical dashed lines indicate the forecast initialization dates relative to the event onset. Forecasts are extended 90 days past the forecast start date.}
\label{fig:extra2016}
\end{figure}


\begin{table}
    \centering
    \small
    \caption{\textbf{Predictability Horizons across the Hikurangi East Coast.} Values represent the mean $\pm$ one standard deviation obtained via temporal convergence analysis across five tectonic regimes (2016–2026).}
    \label{tab:final_mle}
    \begin{tabular}{l c c c}
        \hline
        \textbf{Station} & \textbf{Model (Days)} & \textbf{Raw (Days)} & \textbf{Savgol (Days)} \\
        \hline
        MAHI & $39.0 \pm 3.6$ & $15.5 \pm 1.0$ & $26.0 \pm 3.7$ \\
        GISB & $35.5 \pm 2.6$ & $12.9 \pm 0.3$ & $21.7 \pm 1.3$ \\
        KOKO & $31.9 \pm 2.6$ & $13.9 \pm 0.6$ & $22.5 \pm 1.9$ \\
        CKID & $32.5 \pm 2.7$ & $12.5 \pm 0.4$ & $22.9 \pm 1.6$ \\
        PAWA & $43.8 \pm 6.3$ & $15.0 \pm 0.5$ & $24.5 \pm 1.5$ \\
        CNST & $40.8 \pm 2.8$ & $15.4 \pm 0.6$ & $26.6 \pm 2.3$ \\
        PORA & $40.7 \pm 3.0$ & $12.6 \pm 0.3$ & $22.0 \pm 0.7$ \\
        \hline
        \textbf{Cluster Mean} & \textbf{37.9 $\pm$ 4.3} & \textbf{14.0 $\pm$ 1.2} & \textbf{23.7 $\pm$ 1.9} \\
        \hline
    \end{tabular}
\end{table}

\begin{table}
 \centering
 \small
 \caption{\textbf{GNSS Station Metadata and Spectral Noise Analysis.} Coordinates are retrieved from the GeoNet network. The noise floor ($\sigma_{\text{noise}}$) represents the RMS of residuals following secular detrending and seasonal correction. The spectral index ($\alpha$) characterizes the noise color, where $\alpha \approx 1$ signifies temporally correlated flicker noise.}
 \label{tab:noise_stats}
 \begin{tabular}{lcccc}
  \hline
  \textbf{Station} & \textbf{Lat ($^\circ$S)} & \textbf{Lon ($^\circ$E)} & \textbf{$\sigma_\text{noise}$ (mm)} & \textbf{Index ($\alpha$)} \\
  \hline
  CNST & -38.4880 & 178.2111 & 8.30 & 0.96 \\
  GISB & -38.6353 & 177.8860 & 7.64 & 0.88 \\
  KOKO & -39.0161 & 177.6678 & 5.30 & 0.83 \\
  MAHI & -39.1526 & 177.9070 & 9.08 & 0.86 \\
  CKID & -39.6579 & 177.0764 & 5.09 & 0.83 \\
  PAWA & -40.0331 & 176.8639 & 9.13 & 0.88 \\
  PORA & -40.2664 & 176.6352 & 7.78 & 0.78 \\
  \hline
 \end{tabular}
\end{table}

\begin{table}
 \centering
 \small
 \caption{\textbf{Seismic Triggers and Dynamical Epochs.} Seismic events are collapsed into 180-day dynamical windows to resolve characteristic fault recovery. An 'X' denotes a salient displacement jump ($\ge 2.5$ mm) identified by the automated pipeline. Magnitude $M \ge 7.0$ events (indicated by $^*$) are enforced as regional resets across all stations. Dots ($\cdot$) indicate events that did not meet the saliency threshold for a specific station, and dashes (--) represent periods prior to station installation.}
 \label{tab:event_dates}
 \begin{tabular}{l c ccccccc}
  \hline
  \textbf{Decimal Date} & \textbf{Mag} & \textbf{GISB} & \textbf{KOKO} & \textbf{MAHI} & \textbf{PAWA} & \textbf{CKID} & \textbf{CNST} & \textbf{PORA} \\
  \hline
  2003.0271 & 5.5 & X & -- & -- & -- & -- & -- & -- \\
  2006.3340 & 5.3 & X & X & -- & X & -- & -- & X \\
  2006.9624 & 5.2 & . & . & -- & X & X & -- & X \\
  2007.8977 & 6.6 & X & X & X & . & X & X & X \\
  2008.5873 & 5.5 & . & . & . & . & . & X & . \\
  2009.1452 & 5.2 & . & . & X & . & . & . & . \\
  2009.5666 & 5.4 & . & . & . & . & . & . & X \\
  2010.2053 & 5.1 & . & X & X & X & X & X & . \\
  2010.9389 & 5.6 & . & . & X & . & . & . & . \\
  2011.4161 & 5.5 & . & . & X & . & . & . & X \\
  2011.9700 & 5.8 & . & X & X & . & . & . & . \\
  2012.6343 & 6.3 & . & . & . & X & X & . & X \\
  2013.4926 & 6.5 & . & X & X & . & . & . & . \\
  2015.0049 & 6.7 & X & X & X & . & X & X & . \\
  \textbf{2016.6627}$^*$ & 7.0 & X & X & X & X & X & X & X \\
  \textbf{2016.9458} & 5.9 & X & X & X & X & X & X & X \\
  2017.8247 & 5.5 & . & . & . & . & . & X & . \\
  2018.9955 & 6.1 & . & X & X & . & . & X & . \\
  2019.5520 & 5.3 & . & . & X & . & . & . & . \\
  2020.0354 & 5.8 & . & . & . & . & . & X & . \\
  \textbf{2021.1886}$^*$ & 7.3 & X & X & X & X & X & X & X \\
  2022.9878 & 5.7 & . & X & X & X & X & . & . \\
  2023.3942 & 5.6 & X & X & X & X & . & X & . \\
  2024.9233 & 5.8 & X & X & X & X & . & X & . \\
  \hline
 \end{tabular}
 \begin{flushleft}
  \footnotesize $^*$Regional Reset automatically enforced due to Magnitude $M \ge 7.0$.
 \end{flushleft}
\end{table}

\begin{table}
 \centering
 \small
 \caption{\textbf{Accuracy and Uncertainty Quantification for Physics-Based Ensembles.} Global fidelity is evaluated via the Coefficient of Determination ($R^2$) and RMSE. The model's internal stability is characterized by the ensemble disagreement ($\sigma_{\text{mod}}$), representing the standard deviation across 15 independent trajectories. Total uncertainty ($\sigma_{\text{total}}$) is defined as the quadrature sum of the model disagreement and the instrumental noise floor ($\sigma_{\text{noise}}$). Bolded entries signify high-performing stations used for primary manifold reconstruction and forecasting.}
 \label{tab:fit_accuracy}
 \begin{tabular}{lcccccc}
 \\
 \hline
 \textbf{Station} & \textbf{$R^2$} & \textbf{RMSE (mm)} & \textbf{$\sigma_{noise}$ (mm)} & \textbf{$\sigma_{mod}$ (mm)} & \textbf{$\sigma_{total}$ (mm)} \\
 \hline
 GISB$^\ddagger$ & 0.5375 & 5.193 & 7.64 & 1.868 & 7.865 \\
 KOKO & 0.7401 & 2.700 & 5.30 & 1.720 & 5.572 \\
 \textbf{MAHI} & \textbf{0.8002} & \textbf{4.060} & \textbf{9.08} & \textbf{1.204} & \textbf{9.160} \\
 \textbf{PAWA} & \textbf{0.8742} & \textbf{3.239} & \textbf{9.13} & \textbf{2.785} & \textbf{9.545} \\
 CKID & 0.6106 & 3.178 & 5.09 & 2.370 & 5.615 \\
 \textbf{CNST} & \textbf{0.7960} & \textbf{3.747} & \textbf{8.30} & \textbf{1.828} & \textbf{8.499} \\
 PORA & 0.7529 & 3.867 & 7.78 & 2.882 & 8.297 \\
 \hline
 \end{tabular}
 \begin{flushleft}
 \footnotesize $^\ddagger$Station exhibited significant sensor perturbation and data sparsity during the study period.
 \end{flushleft}
\end{table}

\begin{table}
 \centering
 \small
 \caption{\textbf{Phase Space Diagnostics and False Nearest Neighbors (FNN) Analysis.} Time delays ($\tau$) correspond to the first local minimum of Mutual Average Information (MAI) for physics-based trajectories. FNN percentages represent the proportion of topological self-intersections at an embedding dimension of $m=3$. The physics-based model consistently minimizes false neighbors, justifying the reduction to a 3D manifold. Note: PORA was the only station where the FNN algorithm autonomously identified $m=3$ as the optimal unfolding dimension for the physics-based filter.}
 \label{tab:fnn_percentages}
 \begin{tabular}{l c ccc}
  \hline
  \textbf{Station} & \textbf{$\tau$ (days)} & \textbf{Cleaned Data (\%)} & \textbf{Simple Filter (\%)} & \textbf{Physics Model (\%)} \\
  \hline
  CNST & 81 & 28.72 & 21.90 & 13.90 \\
  GISB & 90 & 27.57 & 20.60 & 14.16 \\
  KOKO & 62 & 33.82 & 23.19 & 17.05 \\
  MAHI & 74 & 28.12 & 19.02 & 14.51 \\
  CKID & 77 & 28.20 & 20.61 & 16.34 \\
  PAWA & 78 & 26.51 & 18.54 & 14.11 \\
  PORA & 88 & 23.87 & 18.69 & 7.54 \\
  \hline
  \textbf{Average} & -- & \textbf{28.12\%} & \textbf{20.36\%} & \textbf{13.94\%} \\
  \hline
 \end{tabular}
\end{table}

\begin{table}
  \centering
  \small
  \caption{\textbf{Optimization Parameter Bounds.} Bounds are defined based on the physical limits of the thermo-poro-mechanical framework and the observed geodetic history of the Hikurangi margin. These constraints ensure the Differential Evolution algorithm converges on a physically consistent manifold.}
  \label{tab:param_bounds}
  \begin{tabular}{llc}
  \hline
  \textbf{Parameter} & \textbf{Physical Meaning} & \textbf{Constraint Range} \\
  \hline
  $\tau$ & Recovery Timescale (years) & [0.1, 3.0] \\
  $r_{scaling}$ & Red Mode Amplitude Scale & [-8.0, 8.0] \\
  $b_{scaling}$ & Black Mode Amplitude Scale & [-8.0, 8.0] \\
  $r_{shift}$ & Red Mode Temporal Shift & [-5.0, 5.0] \\
  $b_{shift}$ & Black Mode Temporal Shift & [-5.0, 5.0] \\
  $v_{drift}$ & Residual Tectonic Velocity (mm/yr) & [15, 25] \\
  $\phi_{init}$ & Initial Modal State & [0.0, 0.25] \\
  \hline
  \end{tabular}
\end{table}

\begin{table}
\centering
\small
\caption{\textbf{Station Distance to Seismic Epicenters.} Distances are in kilometers (km). Bold entries indicate "Far-Field" stations ($>250$ km) where the model's deterministic manifold exhibits increased $\sigma$, correctly signaling spatial decorrelation from the seismic source.}
\label{tab:eventlocations}
\resizebox{\textwidth}{!}{%
\begin{tabular}{l cccc}
\hline 
\textbf{Station} & \textbf{2024 ($M5.8$)} & \textbf{2021 ($M7.3$)} & \textbf{2016 ($M5.9$)} & \textbf{2016 ($M7.0$)} \\
 & 37.64°S, 179.65°E & 37.48°S, 179.46°E & 40.60°S, 177.04°E & 37.36°S, 179.15°E \\
\hline
CNST & 157.8 & 156.6 & \textbf{255.6} & 149.9 \\
GISB & 190.3 & 188.3 & 230.4 & 179.8 \\
KOKO & 231.3 & 231.6 & 184.3 & 225.0 \\
MAHI & 227.0 & 230.1 & 177.4 & 226.9 \\
CKID & \textbf{317.1} & \textbf{318.7} & 105.0 & \textbf{312.7} \\
PAWA & \textbf{359.6} & \textbf{362.3} & 65.0 & \textbf{357.2} \\
PORA & \textbf{391.8} & \textbf{394.6} & 50.6 & \textbf{389.6} \\
\hline 
\end{tabular}%
}
\end{table}

\begin{table}
\centering
\small
\caption{\textbf{Deterministic Manifold Stability.} Forecast stability ($\sigma$, in days) across four major tectonic transients as well as a 2018 'quiet' period for control. The stability $\sigma$ represents the error spread under temporal jittering; lower values indicate a more deterministic, noise-resistant manifold that converges on a singular physical trajectory. Bold values denote the lowest variance ($\sigma$) across the tested methods for a given station and event.}
\label{tab:stability_sigmas}
\resizebox{\textwidth}{!}{%
\begin{tabular}{ll ccccc}
\hline \hline
\textbf{Station} & \textbf{Method} & \textbf{2024 ($M5.8$)} & \textbf{2021 ($M7.3$)} & \textbf{2018 (Quiet)} & \textbf{2016 ($M5.9$)} & \textbf{2016 ($M7.0$)} \\
\hline
\textbf{MAHI} & \textbf{MODEL} & \textbf{0.00} & 1.25 & \textbf{14.39} & \textbf{0.00} & 3.74 \\
& RAW & 1.70 & \textbf{0.82} & 18.17 & 5.44 & 32.90 \\
& SAVGOL & 0.81 & 1.25 & 37.03 & 18.38 & \textbf{3.10} \\
\hline
\textbf{CNST} & \textbf{MODEL} & \textbf{0.47} & \textbf{4.99} & \textbf{2.95} & 2.49 & \textbf{4.55} \\
& RAW & 0.81 & 5.25 & 40.33 & \textbf{2.05} & 7.14 \\
& SAVGOL & 6.95 & 26.00 & 7.59 & 7.14 & 32.56 \\
\hline
\textbf{PAWA} & \textbf{MODEL} & 21.29 & \textbf{10.38} & \textbf{0.82} & \textbf{0.00} & \textbf{0.81} \\
& RAW & 3.74 & 12.37 & 43.16 & 11.57 & 3.30 \\
& SAVGOL & \textbf{1.88} & 31.84 & 1.63 & 1.70 & 16.89 \\
\hline
\textbf{GISB} & \textbf{MODEL} & \textbf{0.47} & \textbf{2.06} & \textbf{0.82} & \textbf{0.81} & \textbf{0.81} \\
& RAW & 0.81 & 4.03 & 3.09 & 1.25 & 13.43 \\
& SAVGOL & 0.81 & 2.95 & 26.21 & 4.93 & \textbf{0.81} \\
\hline
\textbf{KOKO} & \textbf{MODEL} & \textbf{1.26} & \textbf{0.49} & \textbf{1.47} & \textbf{0.00} & \textbf{1.62} \\
& RAW & 5.62 & 8.48 & 3.00 & 10.21 & 34.16 \\
& SAVGOL & 8.52 & 4.22 & 29.18 & 31.88 & 31.04 \\
\hline
\textbf{CKID} & \textbf{MODEL} & \textbf{0.47} & 1.70 & 4.09 & \textbf{0.81} & \textbf{0.81} \\
& RAW & 6.85 & 2.16 & 23.91 & 3.30 & 11.35 \\
& SAVGOL & 43.16 & \textbf{0.82} & \textbf{3.30} & 28.09 & 2.94 \\
\hline
\textbf{PORA} & \textbf{MODEL} & \textbf{0.82} & \textbf{2.16} & \textbf{0.82} & \textbf{0.81} & \textbf{0.81} \\
& RAW & \textbf{0.82} & 3.09 & \textbf{0.82} & 2.05 & \textbf{0.81} \\
& SAVGOL & 17.83 & 4.90 & 3.74 & 7.14 & \textbf{0.81} \\
\hline \hline
\end{tabular}%
}
\end{table}

\clearpage

\end{document}